\documentclass[twocolumn,secnumarabic,amssymb,nobibnotes,aps,prx,superscriptaddress,longbibliography]{revtex4-2}
\usepackage[english]{babel}
\usepackage{graphicx}
\usepackage{amsmath}
\usepackage{amsfonts}
\usepackage[usenames,dvipsnames]{color}
\usepackage{hyperref}
\usepackage{blindtext}
\hypersetup{colorlinks}
\hypersetup{citecolor=blue}
\usepackage[utf8]{inputenc}
\usepackage{xcolor}
\usepackage{mathtools}
\usepackage[normalem]{ulem}
\usepackage{bm}

\def\*#1{\mathbf{#1}}
\def\!#1{\mathbf{\hat#1}}

\begin{document}
\title{ Brownian motion at various length scales with hydrodynamic and direct interactions}
\author{Jeffrey C. Everts}
\email{jeffrey.everts@fuw.edu.pl}
        \affiliation{Institute of Physical Chemistry, Polish Academy of Sciences, 01-224 Warsaw, Poland}
    \affiliation{Institute of Theoretical Physics, Faculty of Physics, University of Warsaw, Pasteura 5, 02-093 Warsaw, Poland}
\author{Robert Hołyst}
        \affiliation{Institute of Physical Chemistry, Polish Academy of Sciences, 01-224 Warsaw, Poland}
\author{Karol Makuch}
\email{kmakuch@ichf.edu.pl}
        \affiliation{Institute of Physical Chemistry, Polish Academy of Sciences, 01-224 Warsaw, Poland}
\date{\today}

\begin{abstract}
Brownian motion is essential for describing diffusion in systems ranging from simple to complex liquids. Unlike simple liquids, which consist of only a solvent, complex liquids, such as colloidal suspensions or the cytoplasm of a cell, are mixtures of various constituents with different shapes and sizes. Describing Brownian motion in such multiscale systems is extremely challenging because direct and many-body hydrodynamic interactions (and their interplay) play a pivotal role. Diffusion of small particles is mainly governed by a low viscous character of the solution, whereas large particles experience a highly viscous flow of the complex liquid on the macro scale. A quantity that encodes hydrodynamics on both length scales is the wavevector-dependent viscosity. Assuming this quantity to be known --in contrast to most studies in which the solvent shear viscosity is given-- provides a new perspective on studying the diffusivity of a tracer, especially in situations where the tracer size can vary by several orders of magnitude. Here, we start systematic studies of exact formal microscopic expressions for the short- and long-time self-diffusion coefficients of a single probe particle in a complex liquid in terms of short-ranged hydrodynamic response kernels.  We study Brownian motion as a function of the probe size, contrasting most theories that focus on self-diffusion as a function of the crowder volume fraction. We discuss the limits of small and large probe sizes for various levels of approximations in our theory, and discuss the current successes and shortcomings of our approach. 
\end{abstract}
\maketitle

\section{Introduction}
The experimental observation of Brownian motion \cite{Brown:1828, Perrin:1909, Huang:2011} and its theoretical understanding \cite{Sutherland:1905, Einstein:1906, Smoluchowski:1906, *[An English translation can be found in:\  ] Cichockibook} is a hallmark in non-equilibrium physics and led to applications ranging from the microscopic understanding of diffusion to the predictability of stock markets \cite{BlackScholes:1973, Merton:1973}. The early work on Brownian motion focused on describing the motion of a single suspended particle (probe) in a medium composed of solvent particles that randomly collide with the surface of the probe particle \cite{Langevin:1908, Uhlenbeck:1930}. For sufficiently long times, the probe enters the diffusive regime, where the translational and rotational diffusion coefficients of a spherical probe are, respectively, given by  \cite{Sutherland:1905, Einstein:1906, Smoluchowski:1906, Debye}
\begin{equation}
D_\mathrm{t,0}=\frac{k_\mathrm{B}T}{6\pi\eta_\mathrm{s}a}, \quad D_\mathrm{r,0}=\frac{k_\mathrm{B}T}{8\pi\eta_\mathrm{s}a^3}. \label{eq:brown}
\end{equation}
Here, $k_\mathrm{B}$ is the Boltzmann constant, $T$ temperature, $\eta_\mathrm{s}$ the shear viscosity of the solvent, and $a$ the particle radius. These expressions -- which can be generalised to the frequency domain \cite{Zia:2018} and more complicated particle shapes \cite{Brenner:1963} -- are widely applied to various systems, sometimes beyond the range of their applicability \cite{Zia:2018}. Indeed, Eq. \eqref{eq:brown} requires strict assumptions. For example, the single-particle picture should be valid (e.g., a suspension at infinite dilution \cite{Krause:2006}), and the probe particle must be macroscopically large compared to the solvent particles. In this case, the solvent acts as a thermal bath, providing a fluctuating force that, together with the kinetic energy of the probe, drives the Brownian motion of the probe \cite{Hinch:1975}. When one of these two assumptions breaks down, more advanced theories are, in principle, needed.

Eq. \eqref{eq:brown} is a valid description for Brownian motion in simple liquids, but breaks down for complex liquids \cite{Hellmann:2011, Holyst:2013, Gupta:2016, Lowen:2020}, such as colloidal suspensions \cite{Pusey:1991, Szamel:1992, Dhont:1992, Pine:1992, Poon:1995, Kalwarczyk:2014}, the cytoplasm of the cell \cite{Roosen:2011, Bubak:2021, Poolman:2022, Huang:2022}, and micellar solutions \cite{Holyst:2006}. Such systems are characterized by at least one other type of particle besides the probe particle and the solvent, which we define as the host particles (or crowders). The reason is an intricate interplay between the direct (i.e., conservative) and hydrodynamic forces between probe and host particles that are not captured within single-particle theory \cite{Hess:1980}. Consequently, on short timescales, the probe is subjected to a force field generated by a fixed configuration of host particles. For longer times, the local environment around a probe particle changes because host particles are also subjected to Brownian motion \cite{Dhont:1996}. Therefore, because of the interactions, a distinction must be made between the short- and long-time self-diffusion coefficients in complex liquids.

For systems that can be carefully experimentally controlled, such as colloidal suspensions, one can explicitly calculate the short- and long-time self-diffusion coefficients with systematic (virial-like) methods for dilute suspensions. Pioneering work by Batchelor \cite{Batchelor:1976, Batchelor:1983,*[corrected in:\ ] Batchelorcorr}, revealed that the short- and long-time translational diffusion coefficient can be expanded around the single-particle result as
\begin{equation}
\frac{D_\mathrm{t,S}}{D_\mathrm{t,0}}=1+\alpha_\mathrm{S}\varphi+\mathcal{O}(\varphi)^2, \quad \frac{D_\mathrm{t,L}}{D_\mathrm{t,0}}=1+(\alpha_\mathrm{S}+\alpha_\mathrm{L})\varphi+\mathcal{O}(\varphi^2), \label{eq:Batchelor}
\end{equation}
with $\varphi$ the volume fraction of crowders based on their hydrodynamic radius. Here, the correction $\alpha_\mathrm{S}$ stems from equilibrium averaged two-body hydrodynamic interactions, whereas distortion of the equilibrium distribution due to Brownian motion of the probe and crowders gives rise to $\alpha_\mathrm{L}$. The $\mathcal{O}(\varphi^2)$ originates from two-body and three-body HIs \cite{Clercx:1992}. Practically, these coefficients can be explicitly computed from the relaxation effect of {\color{black} the crowder-probe pair distribution function in the presence of an external force}, which is easier (but equivalent) than analysing the mean-square displacement for long times \cite{Lekkerkerker:1984}. Eq. \eqref{eq:Batchelor} is valid for arbitrary interaction potentials and has been specifically studied for hard-sphere \cite{Felderhof:1978, Cichocki:1988a, Cichocki:1988}, square-well \cite{vandenbroeck:1985, Cichocki:sw}, and charged fluids \cite{Nagele:1997, Koenderink:2002}, and even extended to arbitrary particle shape and mixtures \cite{Imhof:1995, Koenderink:2001}. Surprisingly, for equal-sized hard spheres, the linear in $\varphi$ contribution is accurate even for dense suspensions, suggesting that at higher orders in $\varphi$, cancellation of various terms occur. Generally, for different-sized hard spheres, such cancellation does not occur \cite{Imhof:1995}. Furthermore, Eq. \eqref{eq:Batchelor} can be straightforwardly extended for rotational Brownian motion, although typically only the short-time regime is considered \cite{Jones:1995}.

For general dense suspensions, various approximate methods have been proposed \cite{Beenakker:1983, Medina:1988, Brady:1994, Peppin:2019}, but often HIs are neglected to make the computations tractable \cite{Dean:2004, Dean:2011, Dhemery:2014}. Furthermore, such methods become increasingly more computationally difficult to handle for multi-component mixtures (e.g., biological fluids) and when the macroscopic effective viscosity of the complex fluid differs by several orders of magnitude from the solvent viscosity (e.g., polymer solutions \cite{Wisniewska:2017}). It should be noted that the HIs mainly cause the latter effect. Thus, a uniform, systematic approach for \emph{all} complex liquids is desirable, which properly accounts for the most important hydrodynamic and direct interactions, while not being restricted to one-component (dilute) colloidal suspensions. Furthermore, most studies focus on diffusion as a function of crowder concentration. In contrast, interesting physics occurs when the probe size is varied (keeping the crowder size constant), especially in the limits of small and large probe sizes. The functional dependence of diffusion as a function of probe size plays a central role in this work.

For a typical (average) hydrodynamic radius of a spherical crowder $R$, Eq. \eqref{eq:brown} is still valid in the short-time regime for $a\ll R$, with $a$ the probe hydrodynamic radius. In the long-time regime, this relation is only approximately true, with a small correction coming from having confined diffusion.
In the opposite limit, $a\gg R$, the expression can still be used provided that $\eta_\mathrm{s}$ is replaced by the effective zero-frequency macroscopic viscosity $\eta_\mathrm{eff}^0$ for long-time diffusion, and the infinite-frequency effective viscosity $\eta_\mathrm{eff}^\infty$ for short-time diffusion. Any good theory should satisfy these limits as a constraint. The interpolation between these two extremes is well described using empirical but physically motivated formulas that fit well to experimental data for various types of complex liquids \cite{Kalwarczyk:2011, Appel:2019} (e.g. obtained with fluorescence correlation spectroscopy \cite{Bonnet:2002, Waxham:2007, Holyst:2009}). However, such expressions are not obtained from first principles; therefore, the connection with microscopic information of the system (e.g., the type of direct interactions) is lacking. 
\begin{figure}[t]
\begin{center}
\includegraphics[width=0.5\textwidth]{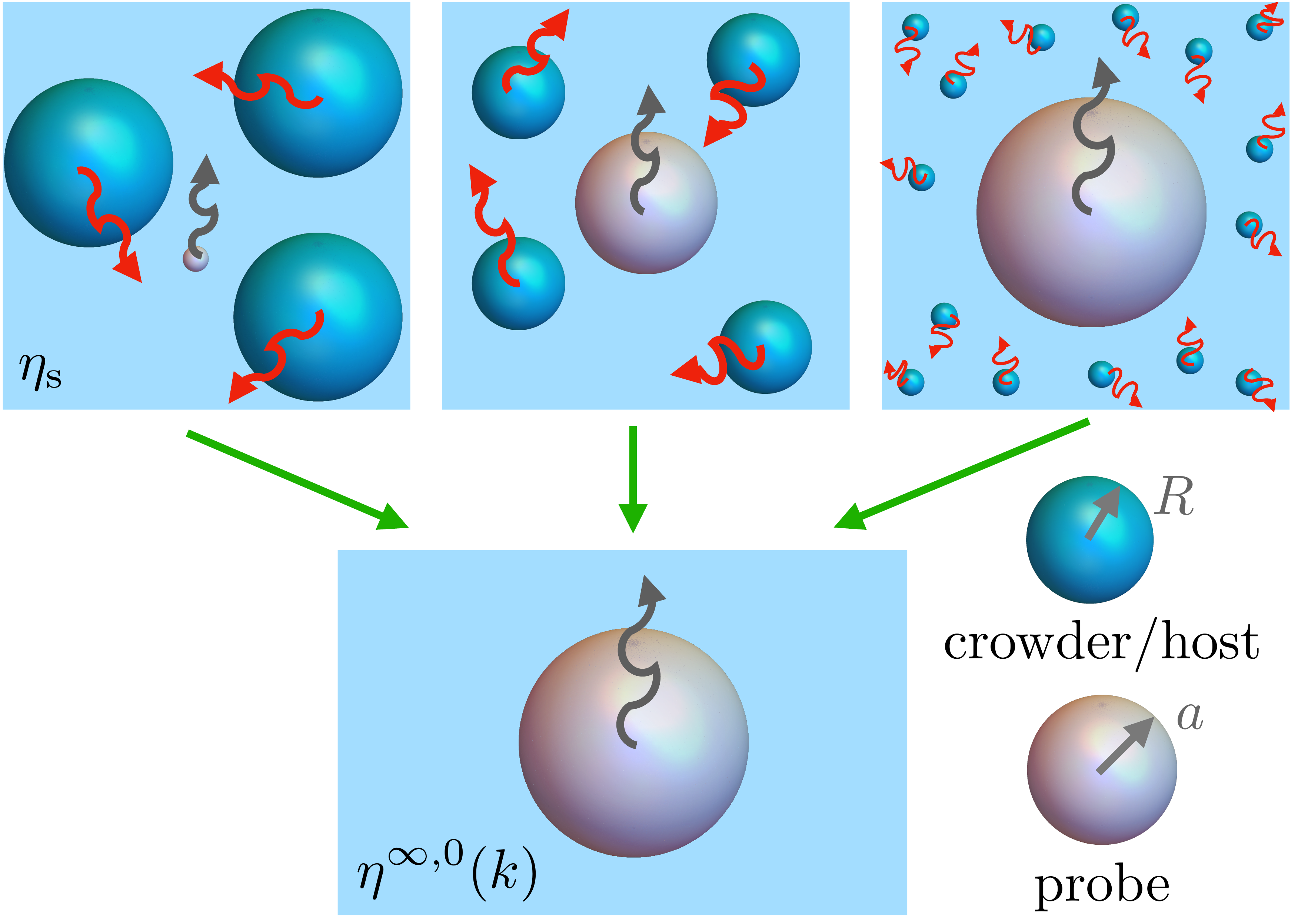} 
\end{center}
\caption{Scheme of solvent picture versus complex-liquid picture for describing self-diffusion. In the solvent picture (first row), self-diffusion is described by the Brownian motion of a tagged probe particle (with radius $a$) in a solvent with shear viscosity $\eta_\mathrm{s}$ and host particles/crowders (with typical radius $R$). In the short-time regime, there is only Brownian motion of the probe particle, and the host particles are stationary (imagine situation without red arrows). In the long-time regime crowders also diffuse. When the probe is varied as a function of size, we hypothesize that $\eta_\mathrm{s}$ is not the proper transport property that should be used to describe diffusion because it only encodes the microscopic length scale of flow in a complex liquid. In the complex-liquid picture, the diffusion is governed by $\eta^\infty(k)$ in the short-time regime, and $\eta^0(k)$ in the long-time regime. These quantities describe flow of the complex liquid on all length scales. } \label{fig:scheme}
\end{figure}

 A study using Batchelor's approach for unequal-sized spheres reveals that the correct limits as a function of $a$ were found for dilute suspensions, see Ref. \cite{Kruger:2009}, and for simplified models of polydisperse systems in Ref. \cite{Miyaguchi:2020}. In this case, the macroscopic effective viscosity satisfies the Einstein formula \cite{Einstein:1906b,*[corrected in:\ ] Einstein:1911}, $\eta_\mathrm{eff}=\eta_\mathrm{s}[1+2.5\varphi+\mathcal{O}(\varphi^2)]$, for the zero-frequency and the infinite-frequency cases \cite{BatchelorGreen:1972,Batchelor:1977}. In Ref. \cite{Kruger:2009}, it was found that $D_\mathrm{t,L}/D_{t,0}$ had a well-defined limit for $a\rightarrow\infty$ only when probe-host HIs were included. Furthermore, the limit converged to the correct value of $\eta_\mathrm{s}/\eta_\mathrm{eff}$ when at least $\mathcal{O}(r^{-8})$ multipoles in the expansion of the mobility tensor were included. This is surprising because the leading order correction to $\eta_\mathrm{s}$ in $\eta_\mathrm{eff}$ can be represented as a one-body contribution, which suggests that the leading order is independent of hydrodynamic and direct interactions. Therefore, we conclude that an expansion around the single-particle solution (Eq. \eqref{eq:Batchelor}) is an inefficient way of computing diffusion coefficients as a function of the probe size for complex liquids where $\eta_\mathrm{eff}^{\infty,0}/\eta_\mathrm{s}\sim 10^1-10^4$, which deviate strongly from a single-particle reference state.

In previous work, an important step towards a description that is valid for all complex liquids has been made by Makuch \emph{et al.} \cite{Makuch:2020}.  The situation is schematically depicted in Fig. \ref{fig:scheme}. Instead of an explicit description of the complex liquid in terms of a solvent described by $\eta_\mathrm{s}$ and crowders, the  microscopic information of the host particles was integrated out to obtain a quantity called the wavevector-dependent viscosity $\eta(k)$, which is experimentally accessible and can be defined for any complex liquid. In Ref. \cite{Makuch:2020}, this idea has been utilized with ad-hoc approximations to describe complex liquids. However, describing diffusion in terms of a given wavevector-dependent viscosity, opens new possibilities for a systematic study of self-diffusion in complex liquids. In particular, the early results of Batchelor can act as an inspiration for such computations. In this paper, we discuss  the role of direct and HIs on the self-diffusion coefficient of a spherical probe particle in a complex liquid, by studying the lowest-order contributions in the short- and long-time diffusive regime. 

This paper is organised as follows. Our starting point is to derive the probability distribution function within linear response theory that we will use to integrate out the host particles (Sec. \ref{sec:prob}). This probability distribution is used to derive the effective Stokes equations in terms of a wavevector-dependent viscosity, describing the flow of a complex liquid on various length scales (Sec. \ref{sec:effstokes}). This quantity needs to be considered in the infinite-frequency $\eta^\infty(k)$ or zero-frequency $\eta^0(k)$ regime depending on the application. From the linear response theory we derive expressions for the short- and long-time self-diffusion coefficients for a given value of $\eta_\mathrm{s}$ (Sec. \ref{sec:solvent}) in terms of the Oseen tensor. Using the effective flow equations, we then derive expressions for the self-diffusion constant of a probe in terms of $\eta^0(k)$ and $\eta^\infty(k)$ -- now assumed to be given quantities-- in the short and long-time regime, respectively (Sec. \ref{sec:complex}). To highlight the potential functionality of our formalism, we will use systematic approximations (Sec. \ref{sec:approx}) to arrive at approximate expressions for the translational and rotational self-diffusion coefficient for a spherical probe in the short-time (Sec. \ref{sec:shorttime}) and long-time regime (Sec. \ref{sec:longtime}). For demonstration purposes, we will evaluate these expressions when the probe only interacts with crowders via steric interactions (Sec. \ref{sec:approximate}). Furthermore, we discuss the limits for $a/R\rightarrow\infty$ and $a/R\rightarrow 0$ for short- and long-time diffusion coefficients. We then summarize our most important results (Sec. \ref{sec:summary}) and discuss our conclusions and the remaining open problems (Sec. \ref{sec:conclusion}).

\section{Probability distribution within linear response theory}
\label{sec:prob}
Consider a complex liquid consisting of (not necessarily identical) host particles of various shapes and sizes immersed in an incompressible solvent. Each host particle is characterized by its centre-of-mass position and internal (rotational and vibrational) degrees of freedom. We model each host particle as a series of interacting spherical beads to simplify notation and not explicitly handle the internal degrees of freedom. For example, a linear polymer chain can be modelled by a bead-spring model \cite{Rouse:1953, Zimm:1956}. The total phase space of the problem can thus be cast in terms of the centre-of-mass positions of beads, ${\bf R}^{N}:=({\bf R}_1,...,{\bf R}_{N})$, with $N$ the total number of beads. 

The probability distribution $P({\bf R}^N,t)$ to find a fixed configuration ${\bf R}^N$ at time $t$ satisfies the continuity equation
\begin{equation}
\frac{\partial}{\partial t}P({\bf R}^N,t)+\sum_{i=1}^N\frac{\partial}{\partial {\bf R}_i}\cdot[{\bf U}_iP({\bf R}^N,t)]=0, \label{eq:cont}
\end{equation}
with ${\bf U}_i$ the Smoluchowski velocity, which only has meaning in the probabilistic sense. For time scales larger than the viscous relaxation time of the solvent, we have that 
\begin{equation}
{\bf U}_i=\sum_{j=1}^N\boldsymbol{\mu}^\mathrm{tt}_{ij}({\bf R}^N)\cdot{\bf F}_j+\int d{\bf r}\, \bm{\mathsf{C}}_i^\mathrm{t}({\bf r;R}^N)\cdot{\bf v}_0({\bf r}),
\end{equation} with $\boldsymbol{\mu}^\mathrm{tt}_{ij}({\bf R}^N)$ the translational-translational mobility tensor and $\bm{\mathsf{C}}_i^\mathrm{t}({\bf r,R}^N)$ the translational convection kernel for stationary viscous flow \cite{Felderhof:1983conv}. Both hydrodynamic quantities are rank-two tensors for which we assume stick boundary conditions on the surfaces of the host particles. Furthermore, ${\bf v}_0$ is the ambient flow field defined as the fluid flow field without host particles ($N=0$). We decompose the total force ${\bf F}_i$ acting on particle $i=1,...,N$ as
\begin{align}
{\bf F}_i&=-\frac{\partial \Phi({\bf R}^N)}{\partial{\bf R}_i}-k_\mathrm{B}T\frac{\partial}{\partial{\bf R}_i}\mathrm{ln} [P({\bf R}^N,t)]+{\bf F}^\mathrm{ext}_i. \label{eq:for}
\end{align}
Here, $\Phi({\bf R}^N)=\sum_{i<j}\phi_{ij}(|{\bf R}_i-{\bf R}_j|)$ is the total interaction potential expressed as a sum of pair potentials. The first term thus describes conservative forces, the second term Brownian forces, and the third term describes the external forces acting on the particles. We are strictly interested in computing the self-diffusion coefficient, for which it is sufficient to consider the steady state probability distribution $P^\infty({\bf R}^N)$ for large times, which is independent of $t$. From Eqs. \eqref{eq:cont}-\eqref{eq:for} it follows that \cite{Deutch:1977, Felderhof:1987b}
\begin{gather}
\mathcal{D}_NP^\infty({\bf R}^N)=\sum_{i=1}^N\frac{\partial}{\partial {\bf R}_i}\cdot\Bigg\{\Bigg[\sum_{j=1}^N\boldsymbol{\mu}^\mathrm{tt}_{ij}({\bf R}^N)\cdot{\bf F}_j^\mathrm{ext}\nonumber \\
+\int d{\bf r}\, \bm{\mathsf{C}}_i^\mathrm{t}({\bf r;R}^N)\cdot{\bf v}_0({\bf r})\Bigg]P^\infty({\bf R}^N)\Bigg\}, \label{eq:statsmo}
\end{gather}
where the $N$-body Smoluchowski differential operator is given by
$$\mathcal{D}_N(...)=\sum_{i,j=1}^N\frac{\partial}{\partial{{\bf R}_i}}\cdot\boldsymbol{\mu}_{ij}^\mathrm{tt}({\bf R}^N)\cdot\Bigg[k_\mathrm{B}T\frac{\partial}{\partial{{\bf R}_j}}+\frac{\partial\Phi({\bf R}^N)}{\partial {\bf R}_j}\Bigg](...).
$$
Eq. \eqref{eq:statsmo} admits a formal solution within linear response theory \cite{Felderhoflin, Felderhof:1983} around the equilibrium distribution $P_\mathrm{eq}({\bf R}^N)=\exp[-\beta\Phi({\bf R}^N)]/Q$ with $Q$ the configurational integral. We find
 \begin{align}
&\frac{P^\infty({\bf R}^N)}{P_\mathrm{eq}({\bf R}^N)}=1+\mathcal{L}_N^{-1}\sum_{i=1}^N\left[\frac{\partial}{\partial{\bf R}_i}-\frac{\partial\beta\Phi({\bf R}^N)}{\partial {\bf R}_j}\right] \label{eq:pinf}\\
&\qquad\cdot\left[\int d{\bf r}\, \bm{\mathsf{C}}_i^\mathrm{t}({\bf r;R}^N)\cdot{\bf v}_0({\bf r})+\sum_{j=1}^N\boldsymbol{\mu}^\mathrm{tt}_{ij}({\bf R}^N)\cdot{\bf F}_j^\mathrm{ext}\right], \nonumber
\end{align}
with the adjoint Smoluchowski operator given by
\begin{equation}
\mathcal{L}_N(...)=\sum_{i,j=1}^N\left[k_\mathrm{B}T\frac{\partial}{\partial{{\bf R}_i}}-\frac{\partial\Phi({\bf R}^N)}{\partial {{\bf R}_i}}\right]\cdot\boldsymbol{\mu}_{ij}^\mathrm{tt}({\bf R}^N)\cdot\frac{\partial}{\partial{{\bf R}_j}}(...)
\end{equation}
and $\mathcal{L}_N^{-1}$ its inverse. See Ref. \cite{Felderhof:1983} for time-dependent generalisations of  Eq. \eqref{eq:pinf}. This manuscript only needs two specific cases of Eq. \eqref{eq:pinf}.  First, inserting ${\bf F}_i^\mathrm{ext}=0$ for all $i$ in Eq. \eqref{eq:pinf} defines the probability distribution $P_\mathrm{V}^\infty({\bf R}^N)$.  
This probability distribution is relevant when no external forces act on the host particles, but there is an ambient fluid flow field. Second, inserting ${\bf v}_0({\bf r})=0$ and ${\bf F}_j^\mathrm{ext}=\delta_{j1}{\bf F}$ in Eq. \eqref{eq:pinf} defines the probability distribution $P_\mathrm{F}^\infty({\bf R}^N)$. In this case, there is no ambient flow, but an external force ${\bf F}$ acts on just one particle, which we define as the probe particle. As we shall see, there is a relation between both averages in the thermodynamic limit.

\section{Effective Stokes equations for flow on various length scales}
\label{sec:effstokes}

Here, we follow the derivation in Ref. \cite{Cichockins}, where the infinite-frequency length-scale dependent flow equations were considered. However, we extend the computation to include also the zero-frequency response. Consider a fixed configuration $\{{\bf R}^N\}$ of host particles in an ambient flow field ${\bf v}_0({\bf r})$, but in the absence of external forces. We define the fluid flow field outside the particles as ${\bf v}_\mathrm{f}({\bf r};{\bf R}^N)$, which we take to be incompressible, $\nabla\cdot{\bf v}_\mathrm{f}({\bf r;R}^N)=0$. In the creeping flow regime, the conservation of linear momentum gives
\begin{gather}
\eta_\mathrm{s}\nabla^2{\bf v}_\mathrm{f}({\bf r;R}^N)-\nabla p({\bf r;R}^N)=0, \label{eq:confStokes}
\end{gather}
where we implicitly assumed time scales much longer than the viscous relaxation time. Eq. \eqref{eq:confStokes} is solved assuming stick boundary conditions on the particle surfaces, with particle velocities ${\bf U}_i+\boldsymbol{\Omega}_i\times({\bf r-R}_i)$ for ${\bf r}$ on the surface of particle $i$. As was shown by Bedeaux and Mazur \cite{Mazur:1974}, Eq. \eqref{eq:confStokes} can be extended to be valid  inside the host particles,
\begin{gather}
\eta_\mathrm{s}\nabla^2{\bf v}({\bf r;R}^N)-\nabla p({\bf r;R}^N)=-{\bf f}_\mathrm{ind}({\bf r,R}^N), \label{eq:indStokes}
\end{gather}
which defines the induced force density ${\bf f}_\mathrm{ind}({\bf r,R}^N)$. For Eq. \eqref{eq:indStokes}, no boundary conditions on the particle surfaces need to be imposed. Furthermore, ${\bf v}({\bf r;R}^N)$ equals ${\bf v}_\mathrm{f}({\bf r;R}^N)$ when ${\bf r}$ is outside any of the host particles and equals ${\bf U}_i+\boldsymbol{\Omega}_i\times({\bf r-R}_i)$ when ${\bf r}$ is inside particle $i$. Therefore, ${\bf v}({\bf r;R}^N)$ is the fluid velocity field for the \emph{entire} complex liquid including the regions inside the particles. The solution to Eq. \eqref{eq:indStokes} is given by
\begin{equation}
{\bf v}({\bf r;R}^N)={\bf v}_0({\bf r})+\int d{\bf r}'\, \bm{\mathsf{G}}({\bf r-r}')\cdot {\bf f}_\mathrm{ind}({\bf r;R}^N), \label{eq:noinsp}
\end{equation}
with $\bm{\mathsf{G}}({\bf r})=({\bf I}+\hat{\bf r}\hat{\bf r})/(8\pi\eta_\mathrm{s}r)$ the Oseen tensor.
The macroscopic flow of the complex liquid in the long-time limit is thus determined by $\langle{\bf v}({\bf r;R}^N)\rangle_\mathrm{V}$, where
\begin{equation}
\langle ...\rangle_\mathrm{V}=\int d{\bf R}^N\, P_\mathrm{V}^\infty({\bf R}^N)(...). \label{eq:pv}
\end{equation}
Our goal is to determine the equations that govern $\langle{\bf v}({\bf r;R}^N)\rangle_\mathrm{V}$. First, we have the incompressibility condition $\nabla\cdot\langle{\bf v}({\bf r;R}^N)\rangle_\mathrm{V}=0$. To determine the equation for the conservation of linear momentum, we express the induced force density as
\begin{gather}
{\bf f}_\mathrm{ind}({\bf r;R}^N)=-\int d{\bf r}'\, \hat{\bm{\mathsf{Z}}}({\bf r,r';R}^N)\cdot{\bf v}_0({\bf r}') \label{eq:find}  \\
+\sum_{i=1}^N\tilde{\bm{\mathsf{C}}}^\mathrm{t}_i({\bf r;R}^N)\cdot\left\{-\frac{\partial \Phi({\bf R}^N)}{\partial{\bf R}_i}-k_\mathrm{B}T\frac{\partial}{\partial{\bf R}_i}\mathrm{ln} [P^\infty_\mathrm{V}({\bf R}^N)]\right\} \nonumber.
\end{gather}
Here, we introduced the convective extended friction kernel $\hat{\bm{\mathsf{Z}}}({\bf r,r';R}^N)$ \cite{Felderhof:1987} and the translational transfer kernel $\tilde{\bm{\mathsf{C}}}^\mathrm{t}_i({\bf r;R}^N)$ \cite{Felderhof:1987}. It can be shown that the transfer kernel and convection kernel are related, i.e. $\tilde{C}_{i,\alpha\beta}^\mathrm{t}=C_{i,\beta\alpha}^\mathrm{t}$ \cite{Felderhof:1987}. Averaging Eq. \eqref{eq:find} using Eq. \eqref{eq:pv}, we find to linear order in ${\bf v}_0({\bf r})$,
\begin{equation}
\langle{\bf f}_\mathrm{ind}({\bf r;R}^N)\rangle_\mathrm{V}=\int d{\bf r}'\,\bm{\mathsf{T}}_\mathrm{V}({\bf r,r'})\cdot{\bf v}_0({\bf r}'), \label{eq:findv}
\end{equation}
with an ambient flow response kernel that can be decomposed as $\bm{\mathsf{T}}_\mathrm{V}({\bf r,r'})=\bm{\mathsf{T}}_\mathrm{V}^\mathrm{ins}({\bf r,r'})+\bm{\mathsf{T}}_\mathrm{V}^\mathrm{ret}({\bf r,r'})$. Therefore, we find that the induced force density has a short-time response due to an ambient flow field governed by the instantaneous response kernel \cite{Szymczak:2004}
\begin{equation}
\bm{\mathsf{T}}_\mathrm{V}^\mathrm{ins}({\bf r,r'})=\langle -\hat{\bm{\mathsf{Z}}}({\bf r,r';R}^N)\rangle_\mathrm{eq}, \label{eq:tvins}
\end{equation}
with
\begin{equation}
\langle ...\rangle_\mathrm{eq}=\int d{\bf R}^N\, P_\mathrm{eq}({\bf R}^N)(...) \label{eq:peq}.
\end{equation}
The order matters in Eq. \eqref{eq:peq} because we will consider averages over quantities containing differential operators.
In particular, the long-time response is governed by the sum of the instantaneous and retarded response kernel, where the latter is given by \cite{Szymczak:2004}
\begin{gather}
\bm{\mathsf{T}}_\mathrm{V}^\mathrm{ret}({\bf r,r'})=k_\mathrm{B}T\Bigg\langle\sum_{i,j=1}^N\tilde{\bm{\mathsf{C}}}_{i}^\mathrm{t}({\bf r};{\bf R}^N)\cdot\frac{\overleftarrow{\partial}}{\partial{\bf R}_i} \nonumber  \\
\mathcal{L}_N^{-1}
 \left[\frac{\partial}{\partial{\bf R}_j}-\frac{\partial\beta\Phi({\bf R}^N)}{\partial {\bf R}_j}\right]\cdot\bm{\mathsf{C}}_j^\mathrm{t}({\bf r}';{\bf R}^N)\Bigg\rangle_\mathrm{eq}. \label{eq:tvret}
\end{gather}
Here, the left-derivative $\overleftarrow\partial_{{\bf R}_i}$ acts on each quantity left of it, which includes $P_\mathrm{eq}({\bf R}^N)$.
In order to derive Eq.~\eqref{eq:tvret}, we used 
\begin{equation}
k_\mathrm{B}T\frac{\partial P_\mathrm{eq}\left({\bf R}^N\right)}{\partial {\bf R}_i}=-\frac{\partial\Phi({\bf R}^N)}{\partial {\bf R}_i}P_\mathrm{eq}\left({\bf R}^N\right) \label{eq:equ}.
\end{equation}
By averaging Eq. \eqref{eq:noinsp}, we eliminate ${\bf v}_0({\bf r})$ from Eq. \eqref{eq:findv} and express $\langle{\bf f}_\mathrm{ind}({\bf r;R}^N)\rangle_\mathrm{V}$ in terms of $\langle{\bf v}({\bf r;R}^N)\rangle_\mathrm{V}$. Inserting this expression in Eq. \eqref{eq:indStokes} results in the integro-differential equation
\begin{gather}
\eta_\mathrm{s}\nabla^2\langle{\bf v(r;R}^N)\rangle_\mathrm{V}-\nabla \langle p({\bf r;R}^N)\rangle_\mathrm{V} \nonumber\\
+\int d{\bf r}'\, \boldsymbol{\Sigma}^0({\bf r,r'})\cdot\langle{\bf v(r';R}^N)\rangle_\mathrm{V}=0, \label{eq:avStokesf}
\end{gather}
where
\begin{equation}
\boldsymbol{\Sigma}^0({\bf r,r'})=\left[(\bm{\mathsf{I}}+\bm{\mathsf{T}}_\mathrm{V}\bm{\mathsf{G}})^{-1}\bm{\mathsf{T}}_\mathrm{V}\right]({\bf r,r'}). 
\label{eq:sigmaa}
\end{equation}
Here, the superscript ``0" denotes a zero-frequency response kernel. Furthermore, we defined the kernel product between two kernels $\bm{\mathsf{A}}({\bf r,r'})$ and $\bm{\mathsf{B}}({\bf r,r'})$ as
$$
[\bm{\mathsf{A}}\bm{\mathsf{B}}]_{\alpha\beta}({\bf r,r}')=\int d{\bf r}''A_{\alpha\lambda}({\bf r,r''})B_{\lambda\beta}({\bf r'',r'}),
$$
with Einstein summation convention implied over Greek indices. Furthermore, we defined the identity kernel as $[\bm{\mathsf{I}}]_{\alpha\beta}({\bf r,r'})=[{\bf I}]_{\alpha\beta}\delta({\bf r-r}')=\delta_{\alpha\beta}\delta({\bf r-r'})$ and inverse $\bm{\mathsf{A}}\bm{\mathsf{A}}^{-1}=\bm{\mathsf{A}}^{-1}\bm{\mathsf{A}}=\bm{\mathsf{I}}$.

Eq. \eqref{eq:avStokesf} can be simplified for translationally invariant and isotropic systems. In this case, $ \boldsymbol{\Sigma}({\bf r,r'})= \boldsymbol{\Sigma}({\bf r-r'})$ and we introduce the Fourier transform as $\tilde{\boldsymbol{\Sigma}}^0({\bf k})=\int d{\bf r}\, \boldsymbol{\Sigma}^0({\bf r})e^{-i{\bf k\cdot r}}$. Eq. \eqref{eq:avStokesf} in Fourier space can then be expressed as
\begin{equation}
-\eta^0(k)k^2\langle{\bf \tilde{v}(k;R}^N)\rangle_\mathrm{V}+i{\bf k}\langle \tilde{p}({\bf k;R}^N)\rangle_\mathrm{V}=0, 
\end{equation}
with the zero-frequency wavevector-dependent viscosity
\begin{equation}
\eta^0(k)=\eta_\mathrm{s}-\frac{1}{2k^2}\mathrm{tr}\left[({\bf I}-\hat{\bf k}\hat{\bf k})\cdot\boldsymbol{\tilde{\Sigma}}^0({\bf k})\right]. \label{eq:etak0}
\end{equation}
The fundamental solution of Eq. \eqref{eq:avStokesf} is, therefore, $\bm{\mathsf{G}}_\mathrm{eff}^0({\bf r,r'})=\left[\bm{\mathsf{G}}(\bm{\mathsf{I}}-\boldsymbol{\Sigma}\bm{\mathsf{G}})^{-1}\right]({\bf r,r'})$, with Fourier transform
\begin{equation}
\tilde{\bm{\mathsf{G}}}_\mathrm{eff}^0({\bf k})=\frac{{\bf I}-\hat{\bf k}\hat{\bf k}}{\eta^0(k)k^2}. \label{eq:geffk}
\end{equation}
Therefore, we have shown that the viscous response of a complex liquid on various length scales is governed by $\eta^0(k)$ at long time scales.

We close this section with a few remarks. Eq. \eqref{eq:etak0} describes the long-time (i.e., zero frequency) wavevector-dependent viscosity. However, the analysis would have been the same for deriving the macroscopic flow equations in the short-time (i.e., infinite-frequency) regime, provided one takes $\bm{\mathsf{T}}_\mathrm{V}=\bm{\mathsf{T}}_\mathrm{V}^\mathrm{ins}$ in Eq. \eqref{eq:sigmaa}, which we define as $\boldsymbol{\Sigma}^\infty({\bf r,r'})$. This defines the infinite-frequency wavevector viscosity
\begin{equation}
\eta^\infty(k)=\eta_\mathrm{s}-\frac{1}{2k^2}\mathrm{tr}\left[({\bf I}-\hat{\bf k}\hat{\bf k})\cdot\boldsymbol{\tilde{\Sigma}}^\infty({\bf k})\right], \label{eq:etak}
\end{equation}
with corresponding effective Green's function defined by the Fourier transform,
\begin{equation}
\tilde{\bm{\mathsf{G}}}_\mathrm{eff}^\infty({\bf k})=\frac{{\bf I}-\hat{\bf k}\hat{\bf k}}{\eta^\infty(k)k^2}. \label{eq:geffk2}
\end{equation}
Generally speaking, if we had used the time-dependent many-body Smoluchowski equation, we would have obtained a macroscopic flow equation governed by a wavevector and frequency-dependent $\eta(k,\omega)$, for which only the special cases of zero-frequency and infinite frequency are relevant in this paper.

Additionally, we notice from Eq. \eqref{eq:etak} that $\eta(k)$ at short length scales ($k\rightarrow\infty$) reduces to the solvent shear viscosity $\eta_\mathrm{s}$, whereas for $k\rightarrow 0$, we define it as the effective viscosity $\eta_\mathrm{eff}^\infty$ governing the flow of a complex liquid on the macroscopic scale at short time scales, and at long time scales by $\eta_\mathrm{eff}^0$. In other words, the macroscopic flow of the complex liquid is described by $\eta_\mathrm{eff}^{\infty,0}\nabla^2\langle{\bf v(r;R}^N)\rangle_\mathrm{V}-\nabla \langle p({\bf r;R}^N)\rangle_\mathrm{V}=0$, as is well known.

\section{Short and long-time self-diffusion coefficients in the solvent picture}
\label{sec:solvent}
To study self-diffusion, we consider a fixed configuration $\{{\bf R}^N\}$ in the absence of an ambient flow field (${\bf v}_0({\bf r})=0$). We select particle $i=1$ to be the probe particle and it is the only particle that is subjected to an external force ${\bf F}$. For this specific setting, the velocity field describing the solid-body motion inside a particle and the fluid flow outside the particle, is
\begin{equation}
{\bf v}({\bf r;R}^N)=\int d{\bf r}'\, \bm{\mathsf{G}}({\bf r-r}')\cdot {\bf f}_\mathrm{ind}({\bf r';R}^N), \label{eq:noinsp2}
\end{equation}
where
\begin{gather}
{\bf f}_\mathrm{ind}({\bf r;R}^N)=\sum_{i=1}^N\tilde{\bm{\mathsf{C}}}^\mathrm{t}_i({\bf r;R}^N)\cdot\Bigg\{-\frac{\partial \Phi({\bf R}^N)}{\partial{\bf R}_i}+\delta_{i1}{\bf F} \nonumber\\
-k_\mathrm{B}T\frac{\partial}{\partial{\bf R}_i}\mathrm{ln} [P^\infty_\mathrm{F}({\bf R}^N)]\Bigg\} \label{eq:find2}.
\end{gather}
Here, $P^\infty_\mathrm{F}({\bf R}^N)$ is defined at the end of Sec. \ref{sec:prob}. Eqs. \eqref{eq:noinsp2} and \eqref{eq:find2} should be contrasted with Eqs. \eqref{eq:noinsp} and \eqref{eq:find}. The fluid velocity field is defined in Eqs. \eqref{eq:noinsp2} only in the probabilistic sense due to the presence of the Brownian force, which is given in the long-time limit by $\langle {\bf v}({\bf r;R}^N)\rangle_\mathrm{F}$.

Because we imposed stick boundary conditions, the probe velocity can be extracted as $\langle{\bf U}\rangle_\mathrm{F}=\langle {\bf v}({\bf R}_1+a\hat{\bf r};{\bf R}^N)\rangle_\mathrm{F}$ and we find to linear order in ${\bf F}$, after shifting integration variables
\begin{equation}
{\bf U}=\left[\int d{\bf r}'\, \bm{\mathsf{G}}(a\hat{\bf r}-{\bf r}')\cdot \bm{\mathsf{T}}_\mathrm{F}({\bf r}')\right]\cdot{\bf F}, \label{eq:almost}
\end{equation}
with the force-response kernel defined via,
\begin{equation}
\bm{\mathsf{T}}_\mathrm{F}({\bf r})\cdot{\bf F}=\langle{\bf f}_\mathrm{ind}({\bf r+R}_1;{\bf R}^N)\rangle_\mathrm{F} \label{eq:findF}.
\end{equation}
The response is governed by the local kernel $\bm{\mathsf{T}}_\mathrm{F}({\bf r})=\bm{\mathsf{T}}_\mathrm{F}^\mathrm{ins}({\bf r})+\bm{\mathsf{T}}^\mathrm{ret}_\mathrm{F}({\bf r})$, where the instantaneous force kernel is
\begin{equation}
\bm{\mathsf{T}}_\mathrm{F}^\mathrm{ins}({\bf r})=\langle\tilde{\bm{\mathsf{C}}}^\mathrm{t}_1({\bf r+R}_1;{\bf R}^N)\rangle_\mathrm{eq}
\end{equation}
and retarded force response kernel is given by \cite{Szymczak:2004}
\begin{gather}
\bm{\mathsf{T}}_\mathrm{F}^\mathrm{ret}({\bf r})
=k_\mathrm{B}T\Bigg\langle\sum_{i,j=1}^N\tilde{\bm{\mathsf{C}}}_{i}^\mathrm{t}({\bf r+R}_1;{\bf R}^N)\cdot\frac{\overleftarrow{\partial}}{\partial{\bf R}_i} \nonumber  \\
\mathcal{L}^{-1}_N
 \left[\frac{\partial}{\partial{\bf R}_j}-\frac{\partial\beta\Phi({\bf R}^N)}{\partial {\bf R}_j}\right]\cdot\boldsymbol{\mu}_{j1}^\mathrm{tt}({\bf R}^N)\Bigg\rangle_\mathrm{eq} \label{eq:whenthisends}
\end{gather}
From symmetry, it follows that the factor between square brackets in Eq. \eqref{eq:almost} is proportional to ${\bf I}$. Together with the fluctuation-dissipation theorem, we thus find that the long-time translational diffusion coefficient is 
\begin{equation}
D_\mathrm{t,L}=\frac{k_\mathrm{B}T}{3}\mathrm{tr}\left[\int d{\bf r}'\, \bm{\mathsf{G}}(a\hat{\bf r}-{\bf r}')\cdot \bm{\mathsf{T}}_\mathrm{F}({\bf r'})\right]. \label{eq:dtl}
\end{equation}
For the short-time translational diffusion coefficient, there is no retarded response,
\begin{equation}
D_\mathrm{t,S}=\frac{k_\mathrm{B}T}{3}\mathrm{tr}\left[\int d{\bf r}'\, \bm{\mathsf{G}}(a\hat{\bf r}-{\bf r}')\cdot \bm{\mathsf{T}}_\mathrm{F}^\mathrm{ins}({\bf r'})\right]. \label{eq:dts}
\end{equation}
We define Eqs. \eqref{eq:dtl} and \eqref{eq:dts} as the solvent picture of computing the translational diffusion coefficient because the expressions depend explicitly on the solvent viscosity $\eta_\mathrm{s}$ through the appearance of $\bm{\mathsf{G}}({\bf r})$. Finally, it should be noted that Eqs. \eqref{eq:dtl} and Eq. \eqref{eq:dts} are equivalent to perhaps more widely used expressions for the short- and long-time translational diffusion coefficients, namely
\begin{equation}
D_\mathrm{t,S}=\frac{k_\mathrm{B}T}{3}\mathrm{tr}\langle\boldsymbol{\mu}^\mathrm{tt}_{11}({\bf R}^N)\rangle_\mathrm{eq} \label{eq:dts2}
\end{equation}
and 
\begin{gather}
D_\mathrm{t,L}=D_\mathrm{t,S}+\frac{k_\mathrm{B}T}{3}\mathrm{tr}\Bigg{\{}k_\mathrm{B}T\Bigg\langle\sum_{i,j=1}^N\boldsymbol{\mu}_{1i}^\mathrm{tt}({\bf R}^N)\cdot\frac{\overleftarrow{\partial}}{\partial{\bf R}_i} \nonumber  \\
\mathcal{L}_N^{-1}
 \left[\frac{\partial}{\partial{\bf R}_j}-\frac{\partial\beta\Phi({\bf R}^N)}{\partial {\bf R}_j}\right]\cdot\boldsymbol{\mu}_{j1}^\mathrm{tt}({\bf R}^N)\Bigg\rangle_\mathrm{eq}\Bigg{\}}. \label{eq:dtl2}
\end{gather}
These formal expressions can be obtained from analysis of the mean-square displacement and the corresponding memory function \cite{Cic:1990}. Furthermore, they are equivalent to Eq. \eqref{eq:Batchelor}, which ultimately stem from linear-response theory combined with a virial-like expansion \cite{Batchelor:1976, Batchelor:1983,*[corrected in:\ ] Batchelorcorr}. 

The same computation can be performed for rotational diffusion. In this paper, we only consider the result for the short-time regime. Now due to the stick boundary conditions, we have $\langle\boldsymbol{\Omega}\rangle_\mathrm{eq}\times(a\hat{\bf r})=\langle{\bf v}({\bf R}_1+a\hat{\bf r})\rangle_\mathrm{eq}$  and we find
\begin{align}
&D_\mathrm{r,S}(a)= \label{eq:dr}\\
&\frac{k_\mathrm{B}T}{2}\mathrm{tr}\Bigg[\hat{\bf r}\cdot\boldsymbol{\epsilon}\cdot
\int d{\bf r}'\, \bm{\mathsf{G}}(a\hat{\bf r}-{\bf r}')\cdot \bm\langle\tilde{\bm{\mathsf{C}}}^\mathrm{r}_1({\bf r}'+{\bf R}_1;{\bf R}^N)\rangle_\mathrm{eq}\Bigg] .\nonumber
\end{align}
with $\boldsymbol{\epsilon}$ the Levi-Cevita tensor and $\tilde{\bm{\mathsf{C}}}^\mathrm{r}_1({\bf r};{\bf R}^N)$ the rotational transfer kernel. We note that Eq. \eqref{eq:dr} is equivalent to
\begin{equation}
D_\mathrm{r,S}=\frac{k_\mathrm{B}T}{3}\mathrm{tr}\langle\boldsymbol{\mu}^\mathrm{rr}_{11}({\bf R}^N)\rangle_\mathrm{eq}  \label{eq:dr2},
\end{equation}
with $\boldsymbol{\mu}^\mathrm{rr}_{ij}({\bf R}^N)$ the rotational-rotational mobility tensor, assuming again stick boundary conditions.

\section{Short and long-time self-diffusion coefficients in the complex-liquid picture}
\label{sec:complex}
We obtained Eqs. \eqref{eq:dtl}, \eqref{eq:dts}, \eqref{eq:dr} in terms of a convolution of the Oseen tensor with a hydrodynamic quantity. We will show that this representation is useful for introducing the effective Greens functions defined in Sec. \ref{sec:prob}, which have Fourier representations Eq. \eqref{eq:geffk} and \eqref{eq:geffk2} depending on the time scale of interest. The general philosophy is that any equilibrium average of a hydrodynamic quantity (e.g., the mobility tensor and the transfer kernel) can be decomposed into short- and long-range parts. The hydrodynamic response kernel with the long-range part subtracted defines the irreducible part of this quantity \cite{Cichocki:2002}. We show that the long-range part can be absorbed in $\bm{\mathsf{G}}({\bf r})$, which naturally leads to expressions of the diffusion coefficient in terms of $\eta^{\infty,0}(k)$. The decomposition of a response kernel in a reducible and irreducible part has been well studied, and we use results discussed in such works \cite{Cichocki:2002, Cichocki:2003, Szymczak:2004, Szymczak:2008}. For example, the transfer kernel of the probe particle can be decomposed as 
\begin{align}
&\langle\tilde{\bm{\mathsf{C}}}^\mathrm{a}_1({\bf r+R}_1;{\bf R}^N)\rangle_\mathrm{eq}= \mathrm{\qquad\qquad\qquad\qquad (a=t,r)}\\
&\int d{\bf r}'\, (\bm{\mathsf{I}}+\langle\hat{\bm{\mathsf{Z}}}_\mathrm{c}\rangle_\mathrm{eq,c}^\mathrm{irr}\bm{\mathsf{G}})^{-1}({\bf r,r'})\langle\tilde{\bm{\mathsf{C}}}^\mathrm{t}_1({\bf r'+R}_1;{\bf R}^N)\rangle_\mathrm{eq}^\mathrm{irr}. \nonumber
\end{align}
Here $\hat{\bm{\mathsf{Z}}}_\mathrm{c}$ is the convective extended friction kernel of the \emph{complex liquid}, i.e.  $\hat{\bm{\mathsf{Z}}}_\mathrm{c}({\bf r,r'})=\hat{\bm{\mathsf{Z}}}({\bf r,r'};{\bf R}_2,...,{\bf R}_N)$. The essential part is that this quantity does not depend on the probe particle, and the equilibrium average of this quantity is over ensembles of the complex liquid without the probe (denoted by ``eq,c"). Furthermore, we have the decomposition,
\begin{equation}
\langle\hat{\bm{\mathsf{Z}}}_\mathrm{c}\rangle_\mathrm{eq,c}^\mathrm{irr}=\langle\hat{\bm{\mathsf{Z}}}_\mathrm{c}\rangle_\mathrm{eq,c}(\bm{\mathsf{I}}-\bm{\mathsf{G}}\langle\hat{\bm{\mathsf{Z}}}_\mathrm{c}\rangle_\mathrm{eq,c})^{-1}.
\end{equation}
In the thermodynamic limit, we identify $\langle\hat{\bm{\mathsf{Z}}}_\mathrm{c}\rangle_\mathrm{eq,c}=\bm{\mathsf{T}}_\mathrm{V}^\mathrm{ins}$ and find an alternative expression of Eq. \eqref{eq:dts},
\begin{equation}
D_\mathrm{t,S}=\frac{k_\mathrm{B}T}{3}\mathrm{tr}\left[\int d{\bf r}'\, \bm{\mathsf{G}}_\mathrm{eff}^\mathrm{\infty}(a\hat{\bf r}-{\bf r}')\cdot \bm{\mathsf{T}}_\mathrm{F}^\mathrm{ins, irr}({\bf r'})\right]. \label{eq:dtcom}
\end{equation}

Using similar decomposition formula for the retarded response kernels, we find in the thermodynamic limit,
\begin{equation}
\bm{\mathsf{T}}_\mathrm{F}({\bf r})=\int d{\bf r}'\, (\bm{\mathsf{I}}-\bm{\mathsf{T}}_\mathrm{V}^\mathrm{irr}\bm{\mathsf{G}})^{-1}({\bf r,r'})\bm{\mathsf{T}}_\mathrm{F}^\mathrm{irr}({\bf r'}) 
\end{equation}
and 
\begin{equation}
\bm{\mathsf{T}}_\mathrm{V}^\mathrm{irr}=\bm{\mathsf{T}}_\mathrm{V}(\bm{\mathsf{I}}-\bm{\mathsf{G}}\bm{\mathsf{T}}_\mathrm{V})^{-1}.
\end{equation}
We stress that  $\bm{\mathsf{T}}_\mathrm{V}$ is the response kernel of the complex liquid in the thermodynamic limit, which does not include the probe.
We conclude that
\begin{equation}
D_\mathrm{t,L}=\frac{k_\mathrm{B}T}{3}\mathrm{tr}\left[\int d{\bf r}'\, \bm{\mathsf{G}}_\mathrm{eff}^0(a\hat{\bf r}-{\bf r}')\cdot \bm{\mathsf{T}}_\mathrm{F}^\mathrm{irr}({\bf r'})\right] \label{eq:dtlcomp}
\end{equation}
and we call this the complex-liquid representation of Eq. \eqref{eq:dtl}. Similarly, the complex-liquid representation of the rotational diffusion coefficient within our model is
\begin{align}
&D_\mathrm{r,S}(a)= \label{eq:drcom} \\
&\frac{k_\mathrm{B}T}{2}\mathrm{tr}\Bigg[\hat{\bf r}\cdot\boldsymbol{\epsilon}\cdot
\int d{\bf r}'\, \bm{\mathsf{G}}_\mathrm{eff}^\mathrm{\infty}(a\hat{\bf r}-{\bf r}')\cdot \bm\langle\tilde{\bm{\mathsf{C}}}^\mathrm{r}_1({\bf r}'+{\bf R}_1;{\bf R}^N)\rangle^\mathrm{irr}_\mathrm{eq}\Bigg],\nonumber
\end{align}
which we only consider in the short-time regime.

\section{Expansions of the transfer kernel and sum rules}
\label{sec:approx}
For computing diffusion coefficients in the complex-liquid picture the hydrodynamic quantities $\boldsymbol{\mu}_{ij}^\mathrm{tt}({\bf R}^N)$ and $\tilde{\bm{\mathsf{C}}}^\mathrm{a}_1({\bf r,R}^N)$ (a=t,r) are the most important. Although mobility tensors have been calculated extensively for various systems, little information is available in the literature on the transfer kernel. In principle, one can compute this quantity from a multiple scattering expansion \cite{Felderhofscat} (e.g., within the multipole picture \cite{FelderhofJ:1989}). For translations, we find for the one-body contribution
\begin{equation}
\tilde{\bm{\mathsf{C}}}^\mathrm{t}_i({\bf r,R}^N)=\frac{\delta(|{\bf r-R}_i|-a_i)}{4\pi a_i^2}+... \label{eq:thisisnotwhatweneed}
\end{equation}
and for rotations
\begin{equation}
\tilde{\bm{\mathsf{C}}}_i^\mathrm{r}({\bf r};{\bf R}^N)=\frac{3}{8\pi a_i^3}\delta(|{\bf r-R}_i|-a_i)\left(\frac{{\bf r-R}_i}{|{\bf r-R}_i|}\cdot\boldsymbol{\epsilon}\cdot{\bf I}\right)+...
\end{equation}
To the best of our knowledge, the higher contributions have never been computed explicitly.

An alternative expansion for the translational transfer kernel can be derived from its symmetry with the translational convection kernel. An expansion of the translational convection kernel due to Felderhof \cite{Felderhof:1987b} can be found from an analysis of a sphere in pure convective motion, i.e., no forces or torques are acting on the sphere. In this case, 
\begin{equation}
{\bf U}_i=\int d{\bf r}\, \bm{\mathsf{C}}_i^\mathrm{t}({\bf r;R}^N)\cdot{\bf v}_0({\bf r}).
\end{equation} 
The form of  $\bm{\mathsf{C}}_i^\mathrm{t}({\bf r;R}^N)$ is not unique: in Ref. \cite{Felderhof:1987b} Felderhof assessed the form of the translational convection kernel in terms of suitable moments of ${\bf v}_0({\bf r})$ defined in the centre of the particle. In this case, we find

\begin{equation}
\bm{\mathsf{C}}_i^\mathrm{t}({\bf r,R}^N)=\delta({\bf r-R}_i){\bf I}-\sum_{j=1}^N\boldsymbol{\mu}_{ij}^\mathrm{td}({\bf R}^N)\cdot\nabla\delta({\bf r-R}_j)+..., \label{eq:felder}
\end{equation}
where higher order terms are proportional to dyads of the form $\nabla...\nabla\delta({\bf r-R}_j)$.
The translational-dipolar mobility tensor $\boldsymbol{\mu}_{ij}^\mathrm{td}$ that appears in Eq. \eqref{eq:felder}, has the following symmetry properties \cite{Cichockihyd:1988},
$$
{\mu}_{ij,\alpha\beta\gamma}^\mathrm{td}={\mu}_{ij,\alpha\gamma\beta}^\mathrm{td}, \quad {\mu}_{ij,\alpha\beta\beta}^\mathrm{td}=0, \quad \mu_{ij,\alpha\beta\gamma}^\mathrm{td}=-\mu_{ji,\beta\gamma\alpha}^\mathrm{dt}.
$$
Together with the property $\tilde{C}_{i,\alpha\beta}^\mathrm{t}=C_{i,\beta\alpha}^\mathrm{t}$, we thus find
\begin{equation}
\tilde{\bm{\mathsf{C}}}_i^\mathrm{t}({\bf r,R}^N)=\delta({\bf r-R}_i){\bf I}+\sum_{j=1}^N\nabla\delta({\bf r-R}_j)\cdot\boldsymbol{\mu}_{ji}^\mathrm{dt}({\bf R}^N)+..., \label{eq:thisiswhatweneed}
\end{equation}
The advantage of Eq. \eqref{eq:thisiswhatweneed} over Eq. \eqref{eq:thisisnotwhatweneed} is that $\boldsymbol{\mu}_{ij}^\mathrm{dt}$ has been computed in the literature \cite{Schmitz:1982, Wajnryb:2013, Zuk:2014, Zuk:2017}. However, there is a subtle point. In the development of the effective Stokes equations (Sec. \ref{sec:effstokes}), we used a $\tilde{\bm{\mathsf{C}}}_i^\mathrm{t}({\bf r,R}^N)$ that results in a fluid velocity field which reproduces the solid-body motion inside of a particle. If we choose the centre of the particle as a reference point (Eq. \eqref{eq:thisiswhatweneed}) we may lose this property. Therefore, Eq. \eqref{eq:thisiswhatweneed} should be considered as an approximation of the expansion Eq. \eqref{eq:thisisnotwhatweneed}. However, Eq. \eqref{eq:thisiswhatweneed} allows for tractable computations that give satisfactory results in various limiting situations, as we shall see. We hypothesize that a minimal model for self-diffusion in complex liquids is thus contained in the expansion of Eq. \eqref{eq:thisiswhatweneed}, which we apply in the remaining sections. It should be noted that the first term in Eq. \eqref{eq:thisiswhatweneed} constitutes the approximation made in Ref. \cite{Makuch:2020} for $\bm{\mathsf{T}}_\mathrm{F}^\mathrm{irr}$ based on empiral grounds. Here, we thus gave an argument for this approximation and provide a minimal extension.

Furthermore, Eq. \eqref{eq:thisiswhatweneed} satisfies exact sum rules for the various response kernels introduced in Sec. \ref{sec:complex}. It follows from Eq. \eqref{eq:thisiswhatweneed} that $\int d{\bf r}\, \tilde{\bm{\mathsf{C}}}_i^\mathrm{t}({\bf r,R}^N)={\bf I}$ and therefore we find the \emph{exact} sum rules
\begin{equation}
\int d{\bf r}\, \bm{\mathsf{T}}_\mathrm{F}^\mathrm{ins}({\bf r})={\bf I}, \quad \int d{\bf r}\, \bm{\mathsf{T}}_\mathrm{F}^\mathrm{ret}({\bf r})=0. \label{eq:sumrule}
\end{equation}
For the latter relation, we used that $-\sum_{i=1}^N\partial_{{\bf R}_i}\Phi({\bf R}^N)=0$ by application of Newton's third law together with pair-wise additivity of $\Phi({\bf R}^N)$. Sum rules for $\tilde{\bm{\mathsf{C}}}_i^\mathrm{r}({\bf r,R}^N)$ are considered in Sec. \ref{sec:rott}.

\section{Explicit computation for short-time diffusion}
\label{sec:shorttime}
The remaining complication for explicit evaluation of Eqs. \eqref{eq:dtcom}, \eqref{eq:dtlcomp}, and \eqref{eq:drcom} is the irreducibility constraint on the response kernels. In this work, we consider only leading-order contributions for which the irreducibility constraint can be ignored; in particular, we mainly focus on one-body contributions from all hydrodynamic quantities in $\bm{\mathsf{T}}_\mathrm{F}^\mathrm{irr}({\bf r})$. Specifically, on the level of transfer kernels, we only keep the first term in the expansion of Eq. \eqref{eq:thisiswhatweneed}.

\subsection{Short-time translational diffusion}
Evaluation of Eq. \eqref{eq:dtcom} with the first term in the expansion of Eq. \eqref{eq:thisiswhatweneed} gives
\begin{equation}
D_\mathrm{t,S}(a)=\frac{k_\mathrm{B}T}{3\pi^2}\int_0^\infty dk\, \frac{j_0(ka)}{\eta^\infty(k)}, \label{eq:dts1}
\end{equation}
which was also derived in Ref. \cite{Makuch:2020}. Note that if instead of Eq. \eqref{eq:thisiswhatweneed} the first term in the expansion of Eq. \eqref{eq:thisisnotwhatweneed} was used, we obtain the same formula as Eq. \eqref{eq:dts1}, but with $j_0(ka)$ replaced by $j_0(ka)^2$. Beenakker found a similar result (see Eq. 9.4 in Ref. \cite{Beenakker:1984}), but with $\eta^\infty(k)$ replaced by the zeroth order contribution in its expansion in terms of correlation functions. However, we hypothesize that Eq. \eqref{eq:dts1} is a better choice because it was shown in Ref. \cite{Makuch:2020} that inversion of this expression gives physical values for the wavevector-dependent viscosity even for complex liquids for which $\eta_\mathrm{eff}^0/\eta_\mathrm{s}$ is large. When expanding the result around the exact one-body transfer kernel (the case with $j_0(ka)^2$), we found unphysical (divergent) values for the wavevector-dependent viscosity. Therefore, it suggests that Eq. \eqref{eq:thisiswhatweneed} is a better starting point than Eq. \eqref{eq:thisisnotwhatweneed}.

At this point, it is good to discuss the limiting behaviour of Eq. \eqref{eq:dts1} as a function of $a$. First, it is clear that in the single-particle limit (only probe+solvent), we retrieve the single-particle result Eq. \eqref{eq:brown}. This result is immediate upon using $\eta^\infty(k)=\eta_\mathrm{s}$ and $\int_0^\infty dk\, j_0(ka)=\pi/(2a)$. Other essential limits for a complex liquid are the small- and large-probe limits. For $a\rightarrow\infty$, we follow the reasoning of Ref. \cite{Beenakker:1984} for approximating integrals of the type in Eq. \eqref{eq:dts1}. For $a$ large, the largest contribution in the integral of Eq. \eqref{eq:dts1} stems from the interval $0<k\leq 1/a$ where $\eta^\infty(k)$ is approximately constant. Therefore, we can approximate the integrand by its zero wavevector, infinite frequency value $\eta_\mathrm{eff}^\infty$. Thus, we find for $a$ large,
\begin{equation}
D_\mathrm{t,S}(a)\approx\frac{k_\mathrm{B}T}{6\pi\eta_\mathrm{eff}^\infty a}, \quad (a\ \mathrm{large}). \label{eq:hqw}
\end{equation}
Note that many multipoles (at least $\mathcal{O}(r^{-8})$) on the level of the mobility tensor were needed in Ref. \cite{Kruger:2009} to obtain this result. In our computation, however, the correct limit is obtained from the first-order term in the expansion of Eq. \eqref{eq:thisiswhatweneed}. This highlights the potential usefulness of computations within the complex-liquid picture because it suggests that just a few multipoles are sufficient. Additionally, we can take the low-volume fraction limit, for which we have $\eta(k)=\eta_\mathrm{s}(1+\varphi f(kR)+....)$ with $f(kR)=5/2-(3/350)(kR)^4+...$\cite{Beenakker:1984}. Taking $k\rightarrow 0$, we conclude that Eq. \eqref{eq:hqw} satisfies also the proper limit for $\varphi$ small. A similar argument exists for the limit of small probes, and we find
\begin{equation}
D_\mathrm{t,S}(a)\approx\frac{k_\mathrm{B}T}{6\pi\eta_\mathrm{s} a}, \quad (a\ \mathrm{small}).
\end{equation}

\subsection{Short-time rotational diffusion}
\label{sec:rott}
Using similar arguments as Felderhof \cite{Felderhof:1987b} for obtaining Eq. \eqref{eq:thisiswhatweneed}, we can derive an expression for the first term in the expansion for the rotational convection kernel. We consider purely convective motion (i.e., particles are force and torque free):
\begin{equation}
\boldsymbol{\Omega}_i=\int d{\bf r}\, \bm{\mathsf{C}}_i^\mathrm{r}({\bf r,R}^N)\cdot{\bf v}_0({\bf r}).
\end{equation}
In the absence of HIs and using the centre of the tagged particles as a reference point, we find that
$\boldsymbol{\Omega}_i=(1/2)[\nabla\times{\bf v}_0({\bf r})]_{{\bf r=R}_i}
$, from which we conclude that $
\bm{\mathsf{C}}_i^\mathrm{r}({\bf r,R}^N)=(1/2)\boldsymbol{\epsilon}\cdot\nabla[\delta({\bf r-R}_i)]+..
$ and therefore by using that $\tilde{C}_{i,\alpha\beta}^\mathrm{r}=C_{i,\beta\alpha}^\mathrm{r}$, we find
\begin{equation}
\tilde{\bm{\mathsf{C}}}_i^\mathrm{r}({\bf r,R}^N)=-\frac{1}{2}\boldsymbol{\epsilon}\cdot\nabla[\delta({\bf r-R}_i)]+...
\end{equation}
Anticipating that higher order terms depend on dyads of $\delta({\bf r-R}_i)$, we find the exact sum rule
\begin{equation}
\int d{\bf r}\, \boldsymbol{\epsilon}\cdot{\bf r}\cdot\tilde{\bm{\mathsf{C}}}_i^\mathrm{r}({\bf r,R}^N)=-{\bf I}.
\end{equation}
For computing diffusion, we insert the first term of this expansion in Eq. \eqref{eq:drcom}, and we find
\begin{equation}
D_\mathrm{r,S}(a)=\frac{k_\mathrm{B}T}{4\pi^2 a}\int_0^\infty dk\, \frac{j_1(ka)k}{\eta^\infty(k)}. \label{eq:rotating}
\end{equation}
In Ref. \cite{Makuch:2020}, the same formula was found on phenomenological grounds by using

\begin{equation}
\tilde{\bm{\mathsf{C}}}_i^\mathrm{r}({\bf r};{\bf R}^N)\approx\frac{3}{8\pi b_i^3}\delta(|{\bf r-R}_i|-b_i)\left(\frac{{\bf r-R}_i}{|{\bf r-R}_i|}\cdot\boldsymbol{\epsilon}\cdot{\bf I}\right),
\end{equation}
with $b_i\rightarrow 0$. We thus support the phenomenological approximation from Ref. \cite{Makuch:2020}.
Using that $\lim_{\lambda\downarrow 0}\int_0^\infty dk\, kj_1(ka)e^{-\lambda k}=\pi/(2a^2)$, we find that Eq.~\eqref{eq:rotating} satisfies the single-particle limit Eq. \eqref{eq:brown}. Furthermore, we find from a similar reasoning as for translations that
\begin{equation}
D_\mathrm{r,S}(a)\approx\frac{k_\mathrm{B}T}{8\pi\eta_\mathrm{eff}^\infty a^3}, \quad (a\ \mathrm{large})
\end{equation}
and
\begin{equation}
D_\mathrm{r,S}(a)\approx\frac{k_\mathrm{B}T}{8\pi\eta_\mathrm{s} a^3}, \quad (a\ \mathrm{small}).
\end{equation}
Inspection of Eq. \eqref{eq:rotating} with Eq. \eqref{eq:dts1} reveals the relation
\begin{equation}
D_\mathrm{r,S}(a)=-\frac{3}{4a}\frac{d}{d a}D_\mathrm{t,S}(a), \label{eq:makuch}
\end{equation}
which has been tested in Ref. \cite{Makuch:2020} to simulation data. Due to the lack of data, the comparison was made for diffusion coefficients acting on different time scales, and therefore a systematic deviation was found.
\section{Explicit computation for long-time translational diffusion}
\label{sec:longtime}
For computation of the long-time translational diffusion, we need to evaluate the irreducible part of Eq.~\eqref{eq:whenthisends}. For simplicity, we assume that all host particles are identical. Because $\eta^0(k)$ contains the effect of host particles to all orders (which do not include the probe), we hypothesize that it is sufficient to approximate $\boldsymbol{\mu}_{ij}^\mathrm{tt}({\bf R}^N)=(6\pi\eta_\mathrm{s}a_i)\delta_{ij}{\bf I}+(N-1)\boldsymbol{\mu}_{ij}^\mathrm{tt}({\bf R}_1,{\bf R}_2)$. We will only discuss two-body contributions to the mobility tensor for which the irreducibility constraint can be dropped. Using this series of approximations, we find a natural decomposition 
\begin{equation}
\bm{\mathsf{T}}_\mathrm{F}^\mathrm{ret,irr}({\bf r})=\bm{\mathsf{T}}_\mathrm{F}^{(1)}({\bf r})+\bm{\mathsf{T}}_\mathrm{F}^{(2)}({\bf r})+...
\end{equation}
with
\begin{gather}
\bm{\mathsf{T}}_\mathrm{F}^{(i)}({\bf r})
=k_\mathrm{B}T(N-1)\left\langle\delta({\bf r}-{\bf R}_{i1})\frac{\overleftarrow{\partial}}{\partial{\bf R}_i}{\bf L}({\bf R}_1,{\bf R}_2)\right\rangle_2 \label{eq:whenthisends2},
\end{gather}
where ${\bf R}_{ij}={\bf R}_i-{\bf R}_j$. The average is defined as
\begin{equation}
\langle ...\rangle_2=\int d{\bf R}_1\int d{\bf R}_2\, P_\mathrm{eq}^{(2)}({\bf R}_1,{\bf R}_2)(...),
\end{equation}
with the  marginal probability distribution 
$$
P_\mathrm{eq}^{(2)}({\bf R}_1,{\bf R}_2)=\int d{\bf R}_3...\int d{\bf R}_N\, P_\mathrm{eq}({\bf R}^N).
$$
Note that the left-derivative in Eq. \eqref{eq:whenthisends2} acts also on $P_\mathrm{eq}^{(2)}({\bf R}_1,{\bf R}_2)$.
The vectorial function ${\bf L}$ is defined by the differential equation
\begin{equation}
\mathcal{L}_2[{\bf L}({\bf R}_1,{\bf R}_2)]=\sum_{j=1}^2\left[\frac{\partial}{\partial{\bf R}_j}-\frac{\partial\beta\phi_{12}(R_{12})}{\partial{\bf R}_j}\right]\cdot\boldsymbol{\mu}_{j1}^\mathrm{tt}({\bf R}_1,{\bf R}_2). \label{eq:diffsomethn}
\end{equation}
The above decomposition shows that the long-time translational diffusion coefficient can be decomposed as
\begin{equation}
D_\mathrm{t,L}=\Delta^{(0)}+\Delta^{(1)}+\Delta^{(2)}.
\end{equation}
where $\Delta^{(1)}$ comes from using the probe transfer kernel $(i=1)$ in Eq. \eqref{eq:whenthisends2} and $\Delta^{(2)}$ from the host transfer kernel  $(i=2)$ in Eq. \eqref{eq:whenthisends2}. The quantity $\Delta^{(0)}$ comes from the instantaneous part of $\bm{\mathsf{T}}_\mathrm{F}^\mathrm{ins,irr}({\bf r})$ in Eq. \eqref{eq:dtlcomp}, and the result is
\begin{equation}
\Delta^{(0)}=\frac{k_\mathrm{B}T}{3\pi^2}\int_0^\infty dk\, \frac{j_0(ka)}{\eta^0(k)}.
\end{equation}
This expression should not be confused with $D_\mathrm{t,S}$, which is governed by $\eta^\infty(k)$, see Eq. \eqref{eq:dts1}. Next, we consider $\Delta^{(1)}$ and $\Delta^{(2)}$, which are the contributions to the retarded response from the probe and host transfer kernel, respectively.

\subsection{Contribution from probe transfer kernel}
\label{sec:idontknow}
We observe that $\bm{\mathsf{T}}_\mathrm{F}^{(1)}({\bf r})\propto\delta({\bf r})$, and straightforward computation gives 
\begin{equation}
\Delta^{(1)}=\frac{k_\mathrm{B}T}{3\pi^2}\int_0^\infty dk\, \frac{j_0(ka)}{\eta^0(k)}\chi(a), \label{eq:Delta1}
\end{equation}
with
\begin{align}
\chi=\frac{\varphi}{R^3} \int_0^\infty d{s}\, s^2g_{12}'(s){\color{black} L}(s). \label{eq:chi}
\end{align}
Here, $\varphi$ is the volume fraction of host particles based on their hydrodynamic radius $R$. Furthermore, we used the translational and rotational invariance (there is no external potential) to write ${\bf L}({\bf R}_1,{\bf R}_2)=\beta {\color{black} L}(|{\bf R}_1-{\bf R}_2|)\hat{\bf R}_{12}$ and $P_\mathrm{eq}^{(2)}({\bf R}_1,{\bf R}_2)=g_{12}(|{\bf R}_1-{\bf R}_2|)/V^2$, where $g_{12}(r)$ is the probe-host radial distribution function {\color{black} of the bulk system.  We conclude that $L(s)$ quantifies to linear order in ${\bf F}$ the distortion of the probe-host pair distribution function in the presense of an external force.} The correction $\Delta^{(1)}$ coincides with the correction computed in the solvent picture when HIs are neglected. However, due to the presence of $\eta^0(k)$, we conclude that $\Delta^{(1)}$ is a correction where probe-host HIs to $D_\mathrm{t,L}$ are neglected {\color{black} on the level of the transfer kernel, but not on the level of mobility, see Eq. \eqref{eq:diffsomethn}}. We call the contributions $\Delta^{(0)}+\Delta^{(1)}$ ``effective single-particle approximation'' (ESP). {\color{black} Note that for comparing the effects of approximate mobility tensors on $\Delta^{(0)}+\Delta^{(1)}$ no explicit form of $\eta^{(0)}(k)$ is needed. Knowledge on the quantity $\chi$ is therefore sufficient since it does not depend on $k$.}

As an additional remark, we see that the integrand in Eq. \eqref{eq:chi} is proportional to $g_{12}'(s)$. From this observation, we conclude that in the computation of $ {\color{black} L}(s)$, we can take the mobility tensor to arbitrary order and omit the irreducibility constraint. The reason is that for two-body contributions, the irreducibility constraint just replaces $g_{12}(r)$ by $h_{12}(r)=g_{12}(r)-1$ \cite{Szymczak:2004}. Both quantities have the same derivative. We will use this property in Sec. \ref{sec:yep}.

\subsection{Contribution from host transfer kernel}
The computation of $\Delta^{(2)}$ is more involved. First, we find an expression for $\bm{\mathsf{T}}_\mathrm{F}^{(2)}({\bf r})$
\begin{align}
\bm{\mathsf{T}}_\mathrm{F}^{(2)}({\bf r})=-\rho\int d{\bf s}\, \Bigg\{\left[\frac{\partial}{\partial {\bf s}}g_{12}(s)\delta({\bf r+s})\right] {\color{black} L}(s)\hat{\bf s}\Bigg\}. 
\end{align}
Partial integration and using that 
\begin{equation}
\frac{\partial}{\partial{\bf s}}[ {\color{black} L}(s)\hat{\bf s}]=\frac{\partial  {\color{black} L}}{\partial s}\hat{\bf s}\hat{\bf s}+\frac{ {\color{black} L}(s)}{s}({\bf I}-\hat{\bf s}\hat{\bf s})
\end{equation}
gives
\begin{align}
\bm{\mathsf{T}}_\mathrm{F}^{(2)}({\bf r})=\rho g_{12}(r)\left[\frac{\partial  {\color{black} L}}{\partial r}\hat{\bf r}\hat{\bf r}+\frac{ {\color{black} L}(r)}{r}({\bf I}-\hat{\bf r}\hat{\bf r})\right].
\end{align}
After some algebra, we find the expression
\begin{align}
&\Delta^{(2)}=\frac{k_\mathrm{B}T}{3\pi^2}\frac{\varphi}{R^3}\int_0^\infty\, dk\, \frac{j_0(ka)}{\eta^0(k)}\int_0^\infty ds\, s^2g_{12}(s) \label{eq:Delta2}
 \\
&\left\{ {\color{black} L}'(s)[j_0(ks)+j_2(ks)]+\frac{ {\color{black} L}(s)}{s}[2j_0(ks)-j_2(ks)]\right\}. \nonumber
\end{align}
The quantities $\Delta^{(1)}$ and $\Delta^{(2)}$ can be explicitly computed when $\phi_{12}(r)$ is specified. We perform such calculations for steric interactions in Sec. \ref{sec:approximate}.

\section{Calculation for steric interactions}
\label{sec:approximate}
\subsection{Single-particle mobility approximation}

First, we consider the case where $\boldsymbol{\mu}_{ij}^\mathrm{tt}=(6\pi\eta_\mathrm{s}a_i)^{-1}\delta_{ij}{\bf I}$ with $a_1=a$ and $a_2=R$. In this case ${\color{black}L}(s)$ satisfies the differential equation

\begin{figure}[t]
\begin{center}
\includegraphics[width=0.45\textwidth]{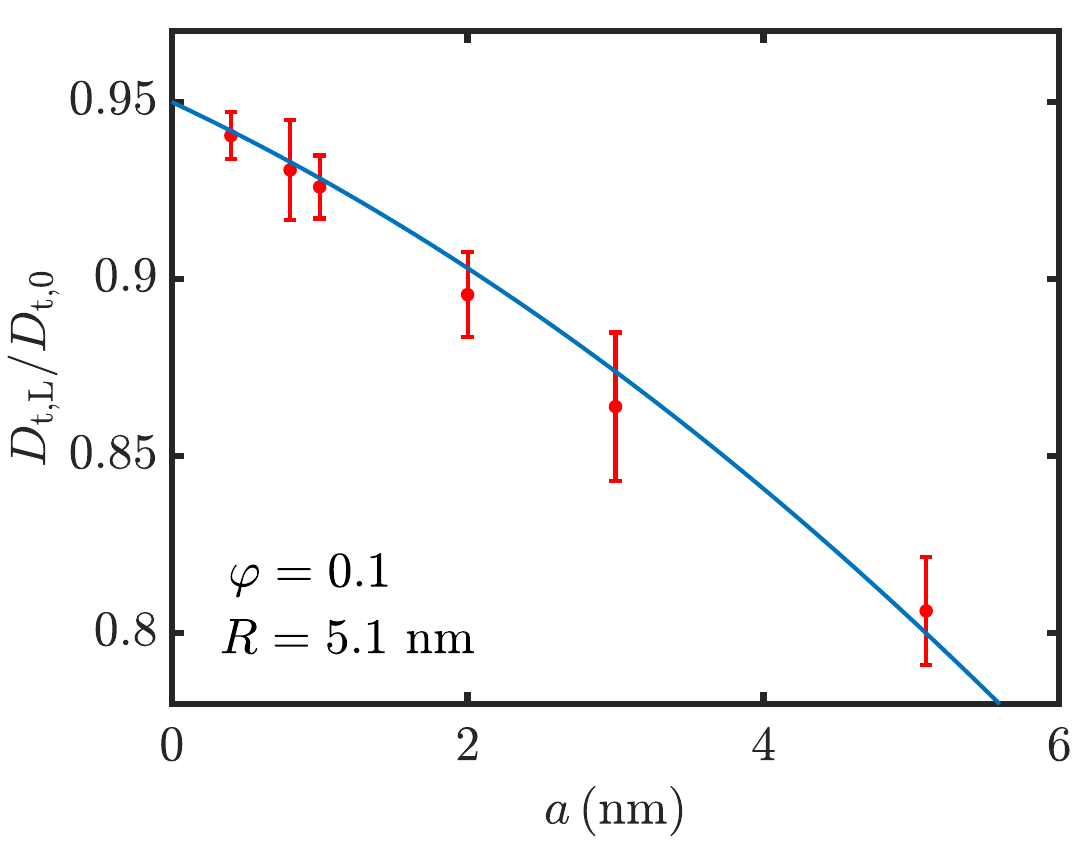} 
\end{center}
\caption{Long-time translational diffusion coefficient scaled to the single-particle result for small probe sizes $a$. The red data points are taken from Brownian dynamics simulations of \cite{Kondrat:2024} without HIs. The full blue line indicates the result from our theory in the $k\rightarrow\infty$ limit without HIs ($\Delta^{(0)}+\Delta^{(1)}$ with single-particle mobility tensor and only probe transfer kernel). 
} \label{fig:small}
\end{figure}

\begin{equation}
\left[\nabla^2   {\color{black} L}(s)-\frac{2  {\color{black} L}(s)}{s^2}\right]- \beta\phi_{12}'(s)  {\color{black} L}'(s)=\frac{R}{a+R} \beta\phi_{12}'(s),
\end{equation}
with a prime denoting differentiation to the argument.
Multiplying this expression with $g_{12}(s)\approx\exp[-\beta\phi_{12}(s)]$, we find
\begin{equation}
\frac{1}{s^2}\frac{d}{d s}\left[s^2g_{12}(s)\frac{d {\color{black} L}}{ds}\right]-\frac{2g_{12}(s) {\color{black} L}(s)}{s^2}=\frac{R}{a+R}g_{12}'(s) \label{eq:superduper}
\end{equation}
to be solved under suitable boundary conditions and specified $\phi_{12}(r)$. For purely steric (additive hard-core) repulsions, we have
\begin{equation}
\phi_{12}(s)=
\begin{cases}
\infty, \quad s<a+R, \\
0, \quad s>a+R.
\end{cases}
\end{equation}
Using that $g_{12}(s)=0$ for $s<a+R$ and $g_{12}(s)=1$ otherwise, and by integrating Eq. \eqref{eq:superduper} from $a+R-\epsilon$ to $a+R+\epsilon$ and letting $\epsilon\downarrow 0$, we find the boundary condition
\begin{equation}
 {\color{black} L}'(x_0^+)=\frac{R}{a+R},
\end{equation}
where $x_0^+=\lim_{\epsilon\downarrow 0}(a+R+\epsilon)$.
Together with the boundary condition $ {\color{black} L}(s\rightarrow\infty)=0$, we find for steric interactions 
\begin{equation}
 {\color{black} L}(s)=-\frac{R}{2}\frac{(a+R)^2}{s^2}, \label{eq:pfunction}
\end{equation}
and from Eqs. \eqref{eq:chi} and \eqref{eq:dtl}
\begin{equation}
\Delta^{(1)}=-\frac{k_\mathrm{B}T}{3\pi^2}\frac{\varphi}{2}\left(1+\frac{a}{R}\right)^2\int_0^\infty dk\, \frac{j_0(ka)}{\eta^0(k)}.\label{eq:dtlhs}
\end{equation}
In Sec. \ref{sec:idontknow} we argued that $\Delta^{(0)}+\Delta^{(1)}$ approximate $D_\mathrm{t,L}$ when probe-host HIs are neglected on the level of the transfer kernel (called ESP), so we identify
\begin{equation}
D_\mathrm{t,L}^\mathrm{ESP}(a)=\frac{k_\mathrm{B}T}{3\pi^2}\int_0^\infty dk\, \frac{j_0(ka)}{\eta^0(k)}\left[1-\frac{\varphi}{2}\left(1+\frac{a}{R}\right)^2\right]. \label{eq:nophHIa}
\end{equation}
For $a$ small, we thus find that
\begin{equation}
\frac{D_\mathrm{t,L}^\mathrm{ESP}}{D_\mathrm{t,0}(a)}\approx1-\frac{\varphi}{2}\left(1+\frac{a}{R}\right)^2+....
\end{equation}
We test this expression to recent simulations (where also host-host HIs are neglected); see Fig. \ref{fig:small}.  We find excellent agreement with the data, including the correct limit $D_\mathrm{t,L}/D_\mathrm{t,0}\rightarrow 1-\varphi/2$ for $a\rightarrow 0$. In Ref. \cite{Kondrat:2024}, the data was fitted using a phenomenological approach. However, here, we find the proper limit from first principles without fit parameters. Furthermore, our computation reveals that the point limit is robust even for high-volume fractions due to the presence of $\eta^0(k)$. This is not obvious within the solvent picture. The opposite limit, $a\rightarrow\infty$, however, reveals that Eq. \eqref{eq:nophHIa} diverges when scaled to $D_\mathrm{t,0}$. This is the same conclusion as Ref. \cite{Kruger:2009}, where no proper macroscopic probe limit was found when probe-host HIs are neglected.

In the next step, we include some of the HIs by computing $\Delta^{(2)}$, assuming a single-particle mobility and one-body host transfer kernel. Using Eq. \eqref{eq:pfunction} in Eq. \eqref{eq:Delta2} gives
\begin{equation}
\Delta^{(2)}=\frac{k_\mathrm{B}T}{\pi^2}\frac{\varphi}{2}\left(1+\frac{a}{R}\right)^2\int_0^\infty dk\, \frac{j_0(ka)}{\eta^0(k)}\frac{j_1(k(a+R))}{k(a+R)}.\label{eq:dtlhs}
\end{equation}
This contribution incorporates part of the host-probe HIs on the level of the transfer kernel because it depends on the positions of probe and host, which might be counterintuitive since it is derivable from the first term in Eq. \eqref{eq:thisiswhatweneed}. To have a consistent treatment of all probe-host HIs, we need to systematically include all two-body effects: (i) the contributions stemming from the second term in Eq. \eqref{eq:thisiswhatweneed} and (ii) terms coming from a two-body mobility tensor which enter Eq. \eqref{eq:diffsomethn}. We focus only on the latter effect, because for the first effect the irreducibility constraint poses many technical difficulties that warrants a deeper analysis.

\subsection{Rotne-Prager-Yamakawa approximation}
\label{sec:yep}
We can restore some of the probe-host HIs by using the full 2-body mobility tensor when solving Eq. \eqref{eq:diffsomethn}. When applied to correct the value for $\Delta^{(1)}$, we can drop the irreducibility constraint as we discussed at the end of Sec. \ref{sec:idontknow}. The minimal, positive definite extension of the single-particle mobility is given by the Rotne-Prager-Yamakawa (RPY) approximation \cite{Rotne:1969, Yamakawa:1970}.  Within this approximation, the self-mobility $\boldsymbol{\mu}_{ii}^\mathrm{tt}$ ($i=1,2$) vanishes, and there is only a non-zero cross mobility
\begin{gather}
(6\pi\eta_\mathrm{s}a)\boldsymbol{\mu}_{12}^\mathrm{tt}({\bf r})= A_{12}^\mathrm{c}(r)\hat{\bf r}\hat{\bf r}\nonumber
+B_{12}^\mathrm{c}(r)({\bf I}-\hat{\bf r}\hat{\bf r}),
\end{gather}
where for $r>a+R$, 
\begin{gather}
A_{12}^\mathrm{c}(r)=\frac{3a}{2r}-\frac{1}{2}a(a^2+R^2)\frac{1}{r^3}+\mathcal{O}(r^{-7}), \nonumber\\
 B_{12}^\mathrm{c}(r)=\frac{3a}{4r}+\frac{1}{4}a(a^2+R^2)\frac{1}{r^3}+\mathcal{O}(r^{-11}).
\end{gather}
Within the RPY approximation, Eq. \eqref{eq:diffsomethn} can be solved with a boundary condition that the probability current vanishes when the two spheres overlap. From the obtained $ {\color{black} L}(s)$, we compute $\chi$ using Eq. \eqref{eq:chi}. We find
\begin{equation}
\chi=-\frac{(a+R)^3(2a+R)(4a+R)(a+4R)\varphi}{(8a^2+43aR+8R^2)(a^4+2a^3R+2aR^3+R^4)}.
\end{equation}
We find for $a\rightarrow 0$ that $\chi\rightarrow 1/2$ is similar to the $\chi$ when a single-particle mobility tensor is assumed. Insertion in Eq. \eqref{eq:Delta1} gives a corrected $\Delta^{(1)}$. The effect of this contribution for small probes is investigated in Fig. \ref{fig:small}, dotted line. Again, the point limit is not affected as compared to our previous results.

For the macroscopic probe limit, we find $\chi\rightarrow-\varphi$ as $a\rightarrow\infty$. We conclude that the obtained $\chi$ parameter gives a well-defined value of $D_\mathrm{t,L}/D_\mathrm{t,0}$ for large probes. However, the value of this limit is
\begin{equation}
\lim_{a\rightarrow\infty}\frac{D_\mathrm{t,L}}{D_{t,0}}=\frac{\eta_\mathrm{s}}{\eta_\mathrm{eff}^0}(1-\varphi), \label{eq:limit2}
\end{equation}
where experimentally, it was found that the limit should (at least approximately) be equal to $\eta_\mathrm{s}/\eta_\mathrm{eff}^0$.

Surprisingly, we find the same limit even when higher-order hydrodynamic multipoles are included for the mobility tensor in the computation of $\Delta^{(1)}$. This is, in a sense, not surprising: the RPY tensor emerges from one reflection in a multiple scattering expansion of the mobility tensor. For consistency, we thus need to include all contributions to $\bm{\mathsf{T}}_\mathrm{F}^\mathrm{ret}({\bf r})$ containing one reflection in the hydrodynamic quantities. Specifically, we have not considered the second term in Eq. \eqref{eq:thisiswhatweneed} which contains $\boldsymbol{\mu}_{ij}^\mathrm{td}$. The result for one reflection of $\boldsymbol{\mu}_{ij}^\mathrm{td}$ has been explicitly computed \cite{Zuk:2014}; however, we need to take the irreducible part when inserting this quantity in the retarded response kernel. Furthermore, the inclusion of such a term will also contribute to the short-time translational diffusion coefficient, which will make $D_\mathrm{t,S}$ explicitly dependent on the interaction potential via the equilibrium distribution.  It would be interesting to consider this contribution for a consistent treatment of probe-host HIs and investigate whether a proper $a\rightarrow\infty$ limit arises. However, this is beyond the scope of the current work.

\subsection{Ad-hoc improvement}
\begin{figure}[t]
\begin{center}
\includegraphics[width=0.45\textwidth]{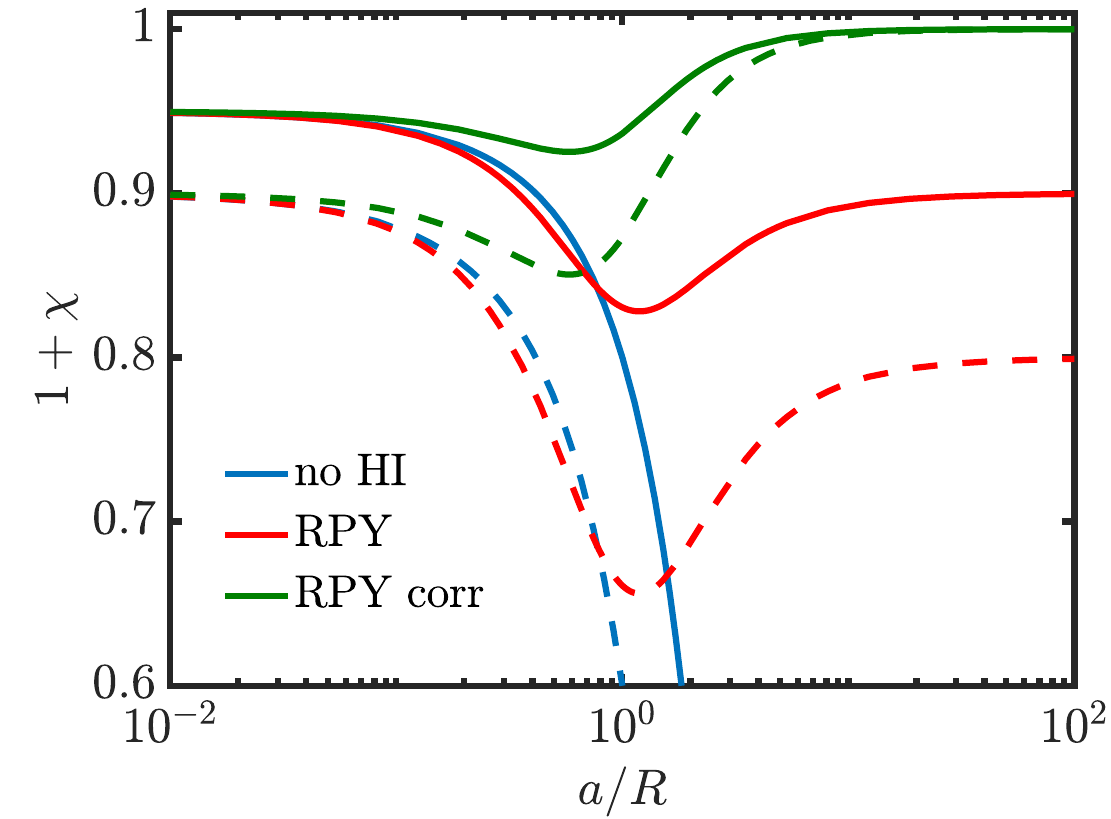} 
\end{center}
\caption{Comparison of various approximations to the quantity $1+\chi$, which is defined as the ratio between the long-time translational diffusion coefficient within the approximation $D_\mathrm{t,L}=\Delta^{(0)}+\Delta^{(1)}$ and $\Delta^{(0)}$ . Here, $\Delta^{(1)}$ is approximated based on how the mobility tensor is treated. For the case ``no HI", we assumed a single-particle mobility tensor, RPY indicates the mobility tensor within the RPY approximation, and ``RPY corr" indicates the proposed result, which is consistent with the low-density result in the long-time regime. The full lines indicate volume fraction $\varphi=0.1$ and the dashed lines $\varphi=0.2$. All three approximations give similar results for small $a/R$, but only the ``RPY corr" case gives the proper macroscopic limit (see main text).} \label{fig:chi}
\end{figure}

One way to remedy the problem of the macroscopic-probe limit is to insist on consistency with the low-$\varphi$ result, which has been computed with Eq. \eqref{eq:Batchelor}. In this calculation, the long-time contribution to translational diffusion $\alpha_\mathrm{L}$ are captured in the so-called Brownian and interaction velocity contributions. Comparison of this method with our calculation reveals that Eq. \eqref{eq:chi} is the same result of the interaction velocity correction in the absence of HIs (in this case, the Brownian velocity vanishes). Within the RPY approximation, the interaction velocity has a different form than Eq. \eqref{eq:chi}, but the Brownian velocity still vanishes. Based on this observation, we propose that within our series of approximations, a suitable candidate for the $\chi$ parameter within the RPY approximation is
\begin{equation}
\chi=\frac{\varphi}{R^3}\int_0^\infty d{r}\,s^2g'(s) {\color{black} L}(s)\left[1-A_{12}^\mathrm{c}(s)\right]. \label{eq:chi2}
\end{equation}
For hard spheres, we then find
\begin{equation}
\chi=-\frac{R^2(2a+R)^2(4a+R)(a+4R)\varphi}{(8a^2+43aR+8R^2)(a^4+2a^3R+2aR^3+R^4)},
\end{equation}
which does satisfy the macroscopic limit $\chi\rightarrow 0$ for $a\rightarrow\infty$. A comparison of the three approximations is given in Fig. \ref{fig:chi} for various approximations on $\chi$. We only performed the comparison up to $\Delta^{(1)}$ for which a comparison on the level of the $\chi$ parameter is sufficient{\color{black}, i.e. no explicit form of $\eta^0(k)$ is needed}. However, for computing $\Delta^{(2)}$ for all probe sizes, $\eta^0(k)$ needs to be specified {\color{black} and is therefore not considered in Fig. \ref{fig:chi}}. The correct macroscopic limit for this ad-hoc improvement of $\Delta^{(1)}$ suggests that a fully consistent treatment of probe-host HIs on the level of one reflection is sufficient to have a correct macroscopic probe limit. Furthermore, it can be seen that all approximations for $\Delta^{(1)}$ give similar results for $a/R$ sufficiently small. 
\section{Summary of main results and open problems}
\label{sec:summary}
The main results of this paper are formally exact expressions for the short- and long-time translational diffusion coefficients in terms of the viscosity functions $\eta^{\infty,0}(k)$, which characterize the length-scale dependent viscous response on different time scales. We assumed in this work that the viscosity function $\eta^{\infty,0}(k)$ is a given quantity. At this point, one could argue that the lowest-order expansion of $\bm{\mathsf{T}}_\mathrm{F}^\mathrm{irr}({\bf r})$ is insufficient for complex liquids, and that at least all two-body terms should be included. We hypothesize that due to the implicit resummation over crowders --that is contained in the viscosity functions $\eta^{\infty,0}(k)$--, that two-body contribution should be sufficient for all hydrodynamic quantities. 

In particular, we found a formula for $D_\mathrm{t,L}$ in the absence of probe-host HIs on the level of the transfer kernel,
\begin{equation}
D_\mathrm{t,L}^\mathrm{ESP}(a)=\frac{k_\mathrm{B}T}{3\pi^2}\int_0^\infty dk\, \frac{j_0(ka)}{\eta^0(k)}\left[1+\chi(a)\right]. \label{eq:nophHI}
\end{equation}
and we analysed the probe-dependent interaction factor $\chi(a)$ for hard spheres. The results compared well with simulation results in the small-probe limit. Furthermore, for the short-time regime we found proper limits in the case of $a\rightarrow 0$ and $a\rightarrow\infty$. The remaining problem lies in the limit $a\rightarrow\infty$, where we were not able to obtain satisfactory results in the long-time regime. We hypothesize that probe-host HIs are essential for this limit, as was already hinted at in previous works by other authors for dilute complex liquids \cite{Kruger:2009}. {\color{black} In this work we have only partially taken into account the probe-host HIs by considering the contributions for which the irreducibility constraint can be dropped.} For future work, it is interesting to systematically analyse {\color{black} all} the probe-host HIs on the level of a two-body approximation {\color{black} to a given (multipole) order} and see whether our hypothesis of a low-density approximation on the level of $\bm{\mathsf{T}}_\mathrm{F}({\bf r})$ is correct.

\section{Conclusion and outlook}
\label{sec:conclusion}
This work underlies the basic framework for computing self-diffusion coefficients in the so-called complex liquid picture where either $\eta^\infty(k)$ (for the short-time regime) or $\eta^0(k)$ (for the long-time regime) are given quantities. In contrast to the picture where the solvent viscosity $\eta_\mathrm{s}$ is given, the complex-liquid picture should be more adequate for describing diffusion in complex liquids at various length scales. Here, the central philosophy is that all hydrodynamic and direct interactions that do not include the probe are effectively resummed and thus contained in $\eta^{\infty,0}(k)$. Here, $\eta^{\infty,0}(k)$ contains information on the viscous flow response on various length scales, which is essential when diffusion is studied as a function of the probe size. Various length scales of the flow are not contained in Eq. \eqref{eq:Batchelor} because the expansion is performed around the single-particle solution, which is governed by the solvent shear viscosity $\eta_\mathrm{s}$. Our results suggest that for the quantity that remains, the irreducible kernel $\bm{\mathsf{T}}_\mathrm{F}^\mathrm{ret,irr}({\bf r})$, only a few terms in the expansion are needed to describe experimentally relevant situations. The trade-off of needing fewer terms in our approach is that two hydrodynamic quantities are required, namely the transfer kernel and the mobility tensor, whereas in established methods only the mobility tensor is needed.

The complex-liquid picture was already introduced in previous work \cite{Makuch:2020}. Here, we started a systematic study of short- and long-time self-diffusion within this framework. We have shown how the quantities $\eta^{\infty,0}(k)$ relate to macroscopic flow so that, indeed, they can be viewed as a length-scale dependent viscous response. Furthermore, we gave support for the phenomenological approximations used in Ref. \cite{Makuch:2020} for $\bm{\mathsf{T}}_\mathrm{F}^\mathrm{irr}({\bf r})$. Here, we used the symmetry between the purely convective motion of particles in ambient flow and the motion of particles under an external force in a quiescent fluid. The zeroth order contribution in the expansion gives approximations similar to the ones in Ref. \cite{Makuch:2020}.

To demonstrate the functionality of our calculation, we have explicitly computed the first terms in the expansion of the diffusion coefficients for additive hard spheres. We obtained satisfactory results for the small-probe limit for translational and rotational diffusion in the short- and long-time regimes. For the large-probe limit, we obtained satisfactory results only in the short-time diffusive regime. For long-time translational diffusion, we hypothesize that including at least one reflection in multiple scattering expansions of hydrodynamic quantities is indispensable to describe probe-host HIs and have a proper macroscopic probe limit for long-time translational diffusion. It is good to realise that elucidating the requirements of achieving such a limit is only possible via theory. Currently, simulations cannot address the large-probe limit due to computational limitations.

Our results open many directions for further work besides finding the minimal level of approximations with a physical macroscopic probe limit. Possible directions include analysing the diffusion coefficient from the mean-squared displacement by direct computation in terms of the wavevector-dependent viscosity. Other directions include investigating the effects of particle shape and activity in the context of biological fluids (for example, the cytoplasm). Finally, an analysis of the effective Stokes equation for a given $\eta^{\infty,0}(k)$ could be an interesting direction for efficiently performing otherwise resource-costly simulations of particles with HIs, like Stokesian dynamics \cite{Brady:1988}. 

\section*{Acknowledgements}
J.C.E. acknowledges financial support from the Polish National Agency for Academic Exchange (NAWA) under the Ulam programme Grant No. PPN/ULM/2019/1/00257. K.M. acknowledges funding from the National Science Centre, Poland
(2021/42/E/ST3/00180). We acknowledge S. Kondrat for useful discussions and for providing us with the raw data of Fig. \ref{fig:small}(a). 
~\newline
\section*{Author declarations}
\subsection*{Conflict of interest}
The authors have no conflicts to disclose.

\subsection*{Data availability}
The data that support the findings of this study are
available from the corresponding author upon reasonable
request.

\bibliography{literature1} 

\end{document}